# Coercive Field Reduction in Ultra-thin $Al_{1-X}Sc_XN$ via Interfacial Engineering with a Scandium Electrode

*AUTHOR NAMES Yinuo Zhang,* [1] *Rajeev Kumar Rai,* [2] *Giovanni Esteves,* [3] *Yubo Wang,*[1] *Deep M. Jariwala,* [1] *Eric A. Stach* [2,]* *and Roy H. Olsson, III* [1,]*

AUTHOR ADDRESS

1. Department of Electrical and Systems Engineering, University of Pennsylvania, Philadelphia, PA, USA 19104
2. Department of Materials Science and Engineering, University of Pennsylvania, Philadelphia, PA, USA 19104
3. Microsystems Engineering, Science and Applications (MESA) Center, Sandia National Laboratories, Albuquerque, NM, USA 87185

* Corresponding Authors: stach@seas.upenn.edu, rolsson@seas.upenn.edu

ABSTRACT

Aluminum scandium nitride (AlScN) ferroelectrics are promising for next-generation non-volatile memory applications due to high remnant polarization compared with $Pb(Zr_xTi_{1-x})O_3$ and doped-$HfO_2$ material systems, as well as their fast switching and scalability to nanometer thicknesses. As AlScN films are thinned to 10 nm thickness, coercive field has been shown to substantially increase, which hinders low voltage operation. We demonstrate that interfacial engineering through bottom electrode selection and strain management reduces this coercive field increase with scaling and improves ferroelectric performance. Here, we demonstrate robust ferroelectricity in ultra-thin AlScN capacitors deposited on a Sc bottom electrode under both alternating current and direct current conditions. The coercive field is reduced by over 20% compared to capacitors with an Al bottom electrode. Furthermore, the difference in dynamic switching behavior over a decade of frequency was evaluated by applying the Kolmogorov–Avrami–Ishibashi (KAI) model. At frequencies <16.7 kHz, the capacitors with Sc and Al bottom electrodes exhibit comparable KAI exponents of 0.030 and 0.028, indicating similar switching kinetics. However, at higher frequencies, the capacitor with an Al bottom electrode shows a significantly higher exponent of 0.063, indicating a stronger frequency dependence, whereas the capacitor with a Sc bottom electrode maintains a stable exponent of 0.030, suggesting a lower frequency dependence during faster switching scenarios. The Scanning Electron Nanobeam Diffraction technique was selected to measure the strain difference in AlScN thin films grown on templates with different lattice mismatch, providing a correlation between lattice mismatch, film strain and switching behavior in ultra-thin film systems.

INTRODUCTION:

Recently, the Sc-alloyed AlN film system has attracted significant interest as a component in next-generation electronics, due to its unique and tunable properties, including ferroelectricity and piezoelectricity.[1-4] Due to its large remnant polarization ($P_r$), low permittivity, and stable wurtzite structure at high temperatures compared to traditional ferroelectric materials,[5-8] the AlScN system has potential applications in non-volatile memory (NVM) technology, including ferroelectric random-access memory (FeRAM) and ferroelectric field-effect transistors (FE-FETs).[9-14] Notably, AlScN film thickness can be reduced to sub-5 nm dimensions and retain ferroelectricity.[15, 16] AlScN is also compatible with Back-End-of-Line (BEOL) Complementary Metal-Oxide-Semiconductor (CMOS) manufacturing.[17, 18] These two features indicate that it can be a key component of multiple third-generation semiconductor commercial applications.[15, 19] Reducing the coercive field to achieve polarization switching will lead to faster switching speed, compatibility with state-of-the-art integrated circuit (IC) transistors, and lower energy consumption, improving the performance of AlScN-based memory devices. Strain engineering has been demonstrated as an effective approach for tailoring the functionality and enhancing the performance of thin films across a range of applications.[20-23] Epitaxial strain at the film interface arises from lattice constant mismatch between the film and the growth substrate.[24-26] This strain can cause crystalline distortions or defect formation that significantly impacts the crystallinity, grain structure, and electronic properties of semiconductor materials.[27] In ferroelectric materials, strain engineering at the interface enables precise modulation of domain structure and key physical properties, including dielectric constants and leakage currents, offering a powerful strategy for tailoring material performance.[27, 28]

In the AlScN system, compressive strain is expected to increase the coercive field ($E_C$) and thus raise the switching energy of memory devices.[4, 29] Increases in the *c/a* ratio close to the interface would therefore be expected to further increase the switching energy barrier. [10, 30] Controlling strain and its relaxation is thus essential to achieve high-quality films in ultra-thin AlScN systems, where device scaling imposes stringent dimensional constraints. However, reducing the coercive field becomes difficult in ultra-thin film systems due to their small crystallite size, which leads to enhanced interface effects. [31-35] Numerous prior studies have shown that the $E_C$ in AlScN sharply increases as the thickness is scaled to 10 nm.[8, 16, 36-38] *Schönweger et al.* demonstrated that the depolarization effect, owing to the finite screening length of electrodes, can partially offset the $E_C$ increase as thickness is further scaled from 10 nm to sub-5 nm. [15, 39] Nevertheless, the selection of growth templates to tailor interfacial strain and crystallinity remains a promising and crucial approach to further improve overall switching performance. [40-43]

In the past several years, various metal films, including {111}-textured Pt and Al, were reported as effective AlScN growth templates to achieve ferroelectric capacitors (FeCaps) and ferroelectric field-effect transistors (FE-FETs).[17, 37, 44-47] However, the strain induced by lattice mismatch in the AlScN unit cell causes a high ferroelectric switching energy barrier in ultra-thin films, resulting in consistently high energy consumption for data storage.[30, 42] *Ryoo et al.* demonstrated that using $HfN_{0.4}$ as a bottom electrode enables the thickness scaling of $Al_{0.7}Sc_{0.3}N$ by maintaining a lattice mismatch ~1% and effectively reducing the interfacial strain.[48] However, they found that $E_C$ could not be reduced for Sc concentrations above 30% due to phase segregation and increased leakage current when the solubility limit was exceeded. Scandium (Sc) oriented along the <002> direction exhibits a hexagonal crystal structure (like AlN) with an *a*-lattice parameter of 3.29 Å compared to $HfN_{0.4}$,[49] which has an *a*-lattice parameter of approximately 3.22 Å.[48] The *a*-lattice parameter of Sc closely matches the *a*-lattice parameter of $Al_{1-x}Sc_xN$, especially for Sc concentrations between 30% to 44% (*a*= 3.25 Å to 3.33 Å).[50-52] This structural compatibility reduces the interfacial strain and allows the incorporation of higher Sc concentrations in the $Al_{1-x}Sc_xN$ system, enabling further reductions in $E_C$. We posit that the utilization of this closer-matched templating layer could facilitate ultra-thin film polarization switching at lower applied electric fields by aiding in the reduction of interfacial strain. We explore this hypothesis explicitly in this report.

In this study, $Al_{0.68}Sc_{0.32}N$ was selected to achieve minimal lattice mismatch with the Sc bottom electrode, leading to reduced $E_C$ values. To investigate the impact of interfacial strain on ferroelectric performance, an ultra-thin AlScN film with the same Sc content was also grown on (111) Al (*a*=2.86 Å).[53] This film exhibited robust ferroelectricity while introducing a distinctly different interfacial strain.[54, 55] This allows a direct comparison of ferroelectric performance of $Al_{0.68}Sc_{0.32}N$ with strain engineered by two different types of bottom electrodes. Due to a smaller lattice mismatch, Sc was grown on GaN/sapphire (*a*-lattice parameter is 3.19 Å for GaN),[56] rather than directly on sapphire (*a*-lattice parameter is 4.785Å),[57] to obtain a highly <002> oriented bottom electrode, whereas Al adopted the strong <111> orientation on sapphire, as reported previously.[37]

To investigate the impact of the bottom electrode on switching dynamics, the kinetic switching behavior of the FeCaps was analyzed using the Kolmogorov–Avrami–Ishibashi (KAI) model.[58] Compared with FeCaps on Al bottom electrodes, devices with Sc bottom electrodes exhibit more stable and reduced frequency dependence of $E_C$ at higher frequencies, indicating robust ferroelectric performance across a broad frequency range. This reduced frequency dependence of $E_C$ enables lower switching voltage and energy consumption under fast switching conditions, making these devices highly promising for high-speed memory applications. [59-62] Scanning

Electron Nanobeam Diffraction (SEND) was employed to further understand the origin of the improved switching behavior.[63] The results showed strain variations in the $Al_{0.68}Sc_{0.32}N$ ultra-thin films arising from differences in the underlying growth templates, offering structural insights into the enhanced ferroelectric performance.

## RESULTS AND DISCUSSION

To verify and evaluate the ferroelectric performance of the AlScN films grown on different templates, we fabricated bottom electrode/$Al_{0.68}Sc_{0.32}N$/Al FeCaps where the metal bottom electrode was either Al or Sc. The device cross-section structure depiction is shown in Figure 1a. The material exhibits Nitrogen-Polarity (N-polar) after deposition and becomes Metal-Polar (M-polar) by applying an electric field larger than the coercive field. The low magnification transmission electron microscopy (TEM) image in Figure 1d shows the thickness of each layer from Sc/$Al_{0.68}Sc_{0.32}N$/Al. The thickness of Al/$Al_{0.68}Sc_{0.32}N$/Al obtained from TEM is shown in Figure S3a. X-ray reflectometry (XRR) spectra (Figure S1) were employed to confirm the thickness of $Al_{0.68}Sc_{0.32}N$ grown on the two growth templates. In this work, we chose the thicknesses obtained from the TEM when calculating the $E_C$ of the FeCaps. This resulted in a slightly higher thickness for the AlScN fabricated on the Al template. This choice leads to the lowest $E_C$ value for the Al/AlScN/Al FeCaps, and a slightly lower thickness for the AlScN fabricated on the Sc template, which leads to the highest $E_C$ value for the Sc/AlScN/Al FeCaps. This selection results in the most conservative analysis when comparing $E_C$ reduction resulting from the better lattice matched Sc template. In Figure 1c, an X-Ray Diffraction (XRD) $\theta$-$2\theta$ scan from 30° to 40° shows that $Al_{0.68}Sc_{0.32}N$ and Sc have a strong $c$-axis out-of-plane orientation. A rocking curve (RC) ω-scan on the Sc 002 reflection generated a Full-Width-Half-Maximum (FWHM) of ~0.76°, indicating the growth of a highly textured Sc film (Figure S2a). In the AlScN deposited on Al, a 002 peak is observed due to the film's low thickness and allows for the *c* lattice parameter to be extracted. The AlScN film deposited on the GaN/Sc template exhibits significant diffuse scattering from the GaN, which complicates the accurate fitting of the AlScN 002 peak. To address this challenge, a triple-axis point detector was used to measure the θ/2θ which effectively suppressed GaN contribution (Figure S2b), resulting in more observable 002 AlScN peak. Due to a higher angular resolution over the PIXcel detector, more accurate *c*-axis parameter detection and rocking curve of the AlScN film can be obtained since it provides a higher angular resolution over the PIXcel detector. RC scans were conducted on both film stacks to obtain the FWHM of 002 AlScN peak. As illustrated in Figure S2c, 3.62°±0.4 and 3.74°±0.23 FWHM of 002 peak were obtained for the AlScN film on Sc and Al bottom electrode, respectively, indicating a similar texture for both samples. The *a*-lattice parameters of Sc and Al were measured by XRD. Based on the 002 and 101 reflections for Sc, the *a*-lattice parameter was calculated to be 3.308Å. From the 111 and 200 reflections of Al, an *a*-lattice parameter of 4.050 Å was determined, corresponding to an equivalent hexagonal close-packed (hcp) *a*-lattice parameter of approximately 2.86 Å. Furthermore, Selected Area Electron Diffraction (SAED) along the AlScN $[11\bar{2}0]$ and Al $[1\bar{1}0]$ zone-axis, as shown in Figure 1e and Figure S3b, demonstrate the *d*-spacing difference of each thin film layer. Energy Dispersive X-ray Spectroscopy (EDS) mapping tracks the elemental composition of the two film stacks (Figure S4 and Figure S5), showing uniform distribution of various elements in their respective layers. Gallium at the bottom electrode/AlScN interface was introduced by ion beam damage during sample preparation. Oxygen was detected on the top Al electrode and Sc layer because of oxidation of the TEM sample when exposed to air. High-magnification TEM images show the interfaces between the ultra-thin $Al_{0.68}Sc_{0.32}N$ film and two types of bottom electrodes (Figure 1f and Figure S3c).

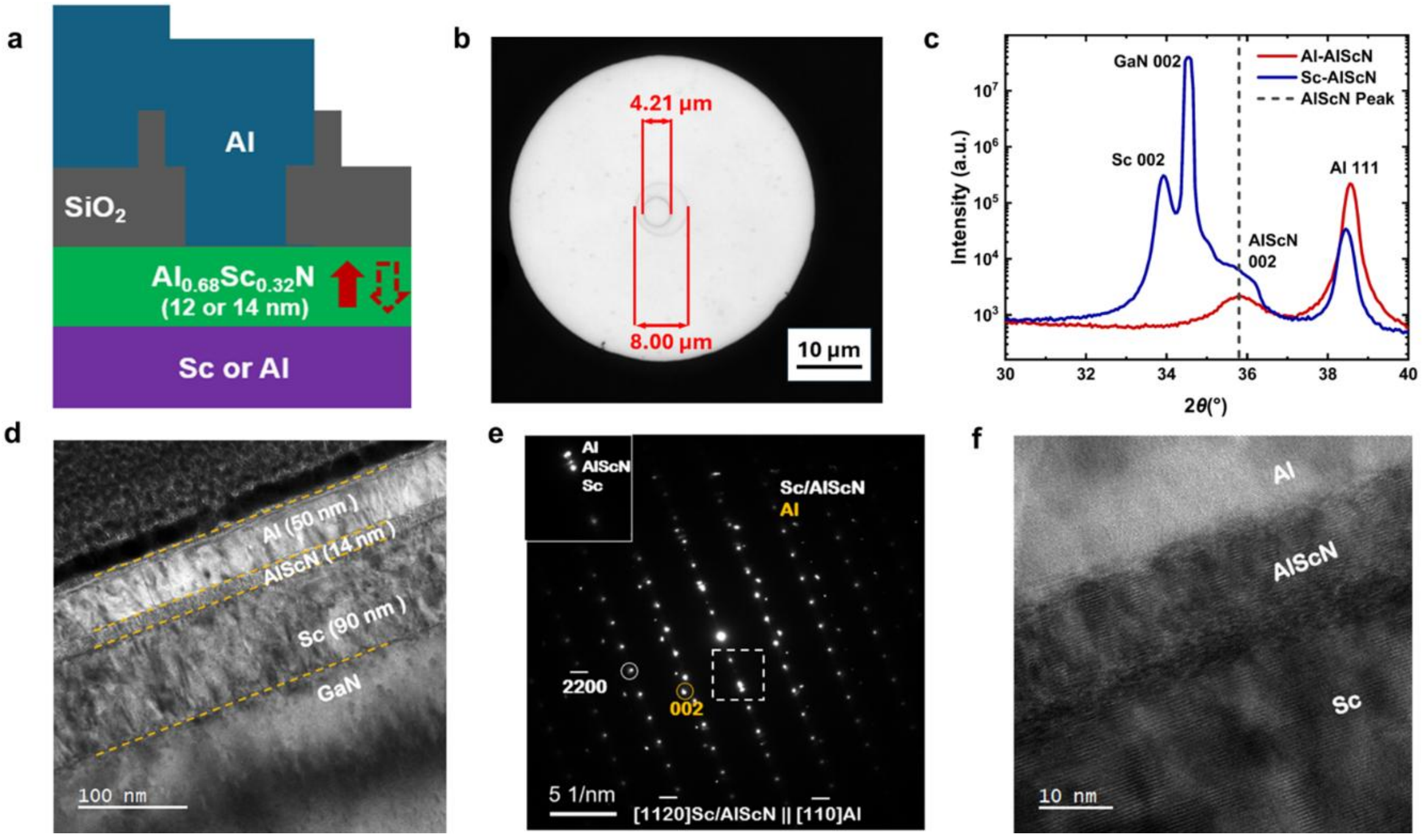

**Figure 1.** Device layout and material structure. (a) Cross-section schematic of $Al_{0.68}Sc_{0.32}N$ ferroelectric capacitors constructed with different bottom electrodes. The dashed arrow in the $Al_{0.68}Sc_{0.32}N$ layer represents the ferroelectric dipole direction after deposition while the solid red arrow represents the direction of the dipole after switching. (b) Optical Microscopy (OM) images of a single FeCap with ~4 µm radius top electrode and ~2.1 µm radius via. (c). $\boldsymbol{\theta/2\theta}$ scan from 30 ° to 40 ° of GaN/Sc/AlScN/Al and Al/AlScN/Al films, showing Sc 002, GaN 002, AlScN 002 and Al 111 peaks. The AlScN peak labeled with a dash line is a peak fitting result of the AlScN 002 for the Al/AlScN/Al device. (d) Cross-sectional low–magnification TEM image showing 90 nm Sc, 14 nm AlScN and 50 nm Al FeCap layers. The interfaces between various layers have been marked by yellow dotted lines. (e) SAED pattern from the Sc/AlScN/Al region along $[\mathbf{11\bar{2}0}]$ of Sc/AlScN and $[\mathbf{1\bar{1}0}]$ of Al. Insert is a magnified section (white dashed boxed region) presenting the out-of-plane spots, showing the different spacings for the three layers (f) High-magnification image of AlScN and the interfaces between Sc/AlScN and AlScN/Al.

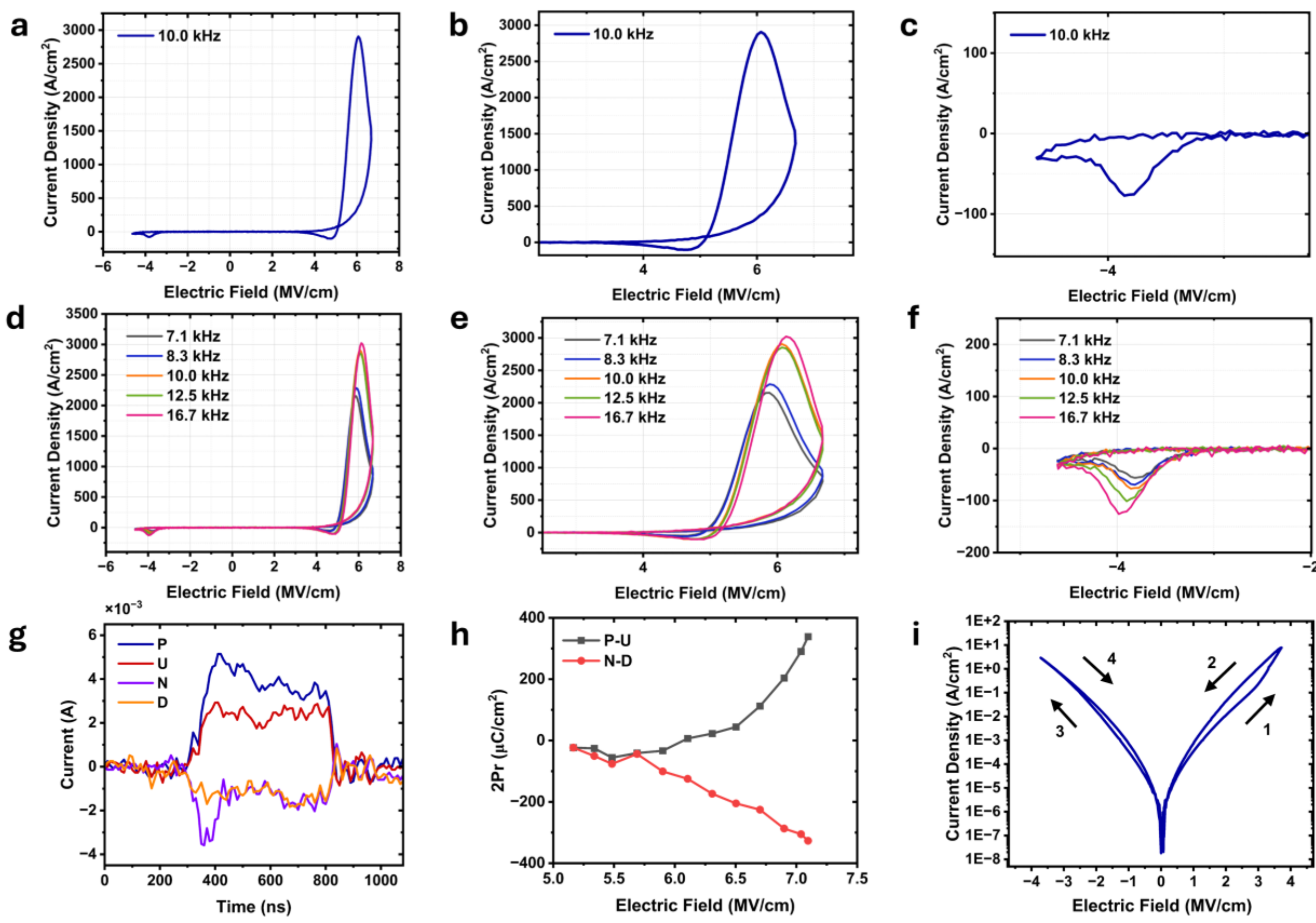

**Figure 2.** Ferroelectric performance of $Sc/Al_{0.68}Sc_{0.32}N/Al$ capacitors. (a) Triangular PUND J-E hysteresis loop under 10 kHz triangular pulses. (b) Magnified image of positive electric-field-induced switching behavior under 10 kHz. (c) Magnified image of negative electric-field-induced switching behavior at 10 kHz. (d) Triangular PUND J-E hysteresis loops under different frequencies. (e) Magnified image of positive electric-field-induced switching performance at different frequencies. (f) Negative electric-field-induced switching comparison at different frequencies. (g) Current-Time response of rectangular PUND measurement with applied electric field of 7.23 MV/cm overlaid in time. The voltage is applied at 9.95 µs with a rise/fall time of 20 ns (h) Polarization change with applied electric field obtained from rectangular PUND measurement. (i) Counterclockwise (CCW) quasi-DC I-V sweep following a sweeping path from 1 to 4.

To study and evaluate the ferroelectric performance of $Sc/Al_{0.68}Sc_{0.32}N/Al$ capacitors, electrical measurements were conducted using a Keithley 4200A-SCS analyzer. AC J-E hysteresis loops, Positive-Up-Negative-Down (PUND) tests, and DC I-V sweeps were used to analyze the switching behavior of the films. A 4.5 kHz triangular waveform (Figure S6(a)) was applied to the device over a specified electric field range from 6.7 MV/cm to -4.6 MV/cm. In Figure S6(b), both positive coercive field ($E_C^+$) and negative coercive field ($E_C^-$) can be clearly observed and are determined to be 5.93 MV/cm and -4.02 MV/cm, respectively. The asymmetry in the coercive field between positive and negative sweeps is driven by the imprint effect, which has been frequently observed in sputtered AlScN films.[36, 64] The switching and leakage behaviors were investigated by varying sweeping frequencies, as illustrated in Figure S7. Ferroelectric switching requires higher electric fields at higher frequencies due to the reduced dipole reorientation time and pinning effects, hindering domain motion and increasing the switching barrier.[65-69] As the frequency increases, the negative coercive fields become more pronounced (Figure S7(c)), whereas the ferroelectric switching at positive electric fields gradually merges with the resistive leakage peak (Figure S7(b)), making its precise identification more challenging.

To clearly observe the $E_C$, a triangular PUND technique previously applied to identify the ferroelectricity in AlScN thin film FeCaps under elevated temperatures was employed (Figure S8(a)).[70] Two sets of triangular I-V pulses were utilized. The response to the first pulse shows both ferroelectric switching and leakage currents, while the response to the second pulse measured only leakage. The compensated current density peak was obtained by subtracting the first peak from the second (Figure S8(b)). In Figure 2a-2c, a triangular PUND test of the capacitor at 10 kHz excitation is presented. It shows pronounced ferroelectric switching peaks. The $E_C^+$ and $E_C^-$ could be clearly identified after leakage subtraction as 6.06 MV/cm and -3.86 MV/cm, respectively. This technique was further applied across frequencies. In Figure 2d, a sweep range from 6.7 MV/cm to -4.6 MV/cm was maintained to ensure consistency with the single J-E hysteresis loop setup. Notably, switching events became more clearly defined after leakage subtraction for both positive and negative applied fields. Although leakage compensation has been employed to determine the $E_C$, the displacement currents for positive and negative applied fields showed a strong asymmetry, which is discussed in Note 1 of the SI.

To further detect the dipole switching while gradually increasing the applied electric field, rectangular PUND measurements with 500 ns pulses were conducted. Reducing the pulse duration reduces the impact from resistive leakage for ultra-thin film AlScN materials,[37, 71] enabling more accurate measurement of the switching polarization ($2P_r$). The $2P_r$ is found by subtracting the polarization measured during the "U" and "D" pulses from that measured during the "P" and "N" pulses. Using 500 ns rectangular pulses (Figure S9(a)), clear dipole switching could be observed via current time response under an applied voltage of 10.13 V, as shown in Figure S9 (b), and after time overlaying the responses to each pulse as shown in Figure 2g. The switching from N-polar to M-polar using positive applied electric fields exhibited a slower switching response than when switching from M-polar to N-polar using negative applied electric fields. As shown in Figure 2h, polarization switching was first observed at ∼5.7 MV/cm. The polarization increased with applied electric fields, reflecting the progressive switching process. However, precise determination of the saturated polarization remained challenging due to the high leakage observed in the ultra-thin film.[72-74]

Next, quasi-DC I-V measurements were conducted to investigate switching performance under slow sweeping conditions. A DC I-V hysteresis measurement was performed on the capacitor, sweeping from 5.2 V to -5.2 V with a 0.025 V step size. In Figure 2i, counterclockwise (CCW) current density sweeping behavior was observed, with higher current density along paths 2 and 4. To examine the electric-field-induced dipole switching, the first derivative of current with respect to voltage vs. the applied field was plotted (Figure S10). The peaks in the $\frac{di}{dV}$ response at 3.30 MV/cm and -3.28 MV/cm correspond to the ferroelectric dipole switching, which caused an abrupt increase in current. The measurement exhibited asymmetric sweeping consistent with the previously reported ferrodiode effect in AlScN. [1, 75]

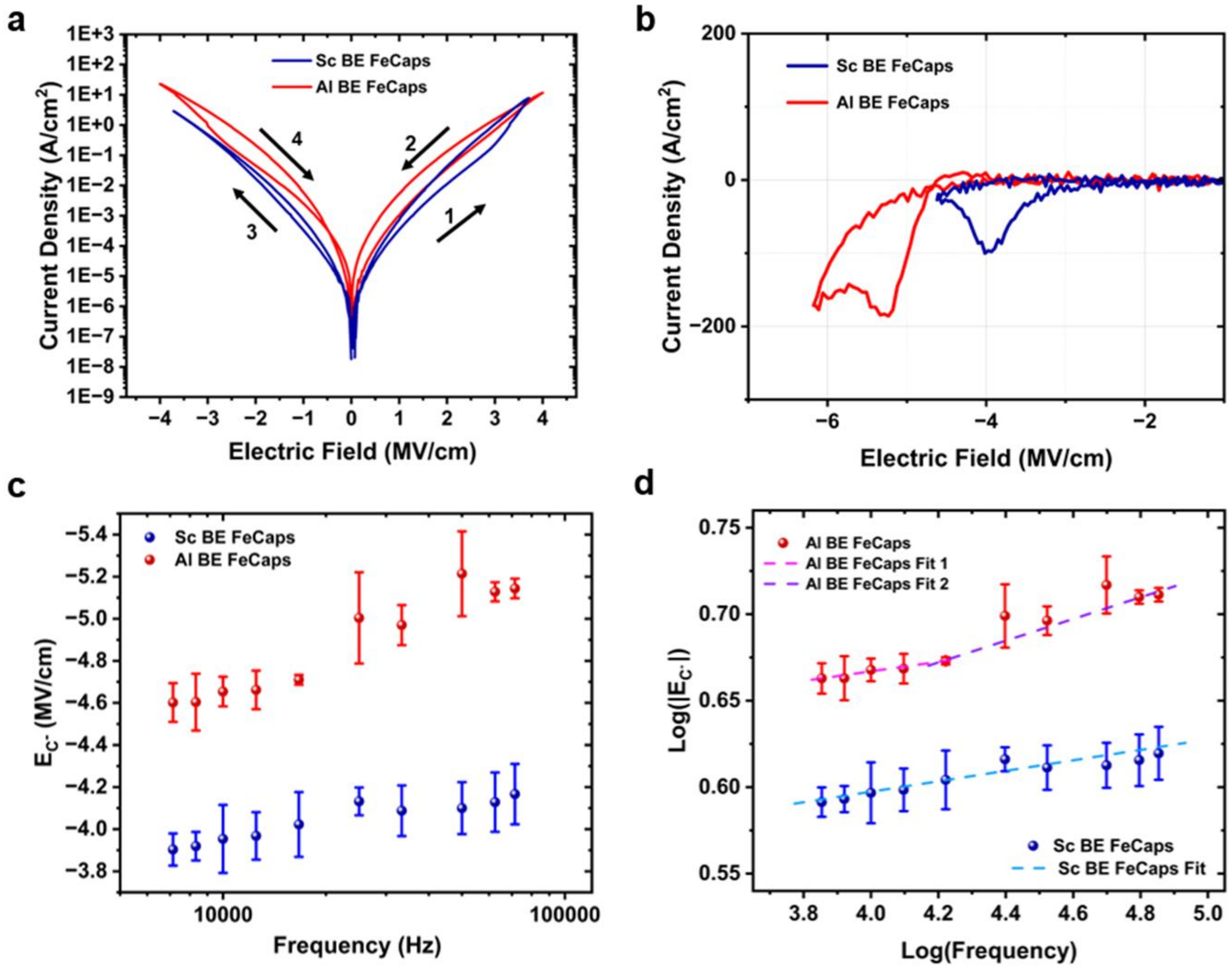

**Figure 3.** Comparison of the ferroelectricity of AlScN on Sc and Al bottom electrodes. (a) Quasi-DC I-V current density vs. applied electric field. (b) $E_C^-$ comparison between Sc/$Al_{0.68}Sc_{0.32}N$/Al capacitors and Al/$Al_{0.68}Sc_{0.32}N$/Al capacitors using 25 kHz Triangular PUND J-E hysteresis loops. (c) $E_C^-$ with error bars versus frequency for the capacitors with different bottom electrodes. (d) Logarithmic power-law fitting of $|E_C^-|$ vs. frequency for Sc BE FeCaps and Al BE FeCaps over a frequency range from 7.1 kHz to 71 kHz.

The ferroelectric behavior of $Al_{0.68}Sc_{0.32}N$ grown on Sc and Al bottom electrodes is compared by focusing on their switching performance and frequency dependence using DC and AC electrical measurement conditions, as shown in Figure 3. In Figure 3a, the CCW DC I-V sweep of both devices – which tracks the change in current density with the applied electric field – shows that FeCaps based on AlScN with an Al electrode exhibit higher leakage current before the onset of polarization switching. The $E_C$ for positive and negative switching has been determined by the 1st derivative of the current with respect to voltage, $\frac{di}{dV}$, versus electric field (Figure S11). The measured $E_C$ for positive and negative switching on FeCaps grown on an Al bottom electrode (Al BE) are 3.92 MV/cm and -3.67 MV/cm, respectively. Compared with Al BE FeCaps, those with Sc bottom electrode (Sc BE) show reduced $E_C$ by 15.8% ($E_C^+$) and 10.6% ($E_C^-$).

Figure 3b shows the switching performance under AC I-V conditions. A triangular PUND pulse test at 25 kHz was conducted on Sc BE FeCaps and Al BE FeCaps, indicating an $E_C^-$ of -4.02 MV/cm and -5.23 MV/cm, respectively. This corresponds to a ~23.1% reduction in $E_C^-$ for the FeCaps grown on Sc. FeCaps grown on an Al BE exhibit more substantial resistive leakage at negative applied electric field, consistent with the behavior observed in the DC I-V sweeps. The higher applied electric field (8.80 MV/cm to -6.17 MV/cm) in FeCaps on Al electrodes, compared to the lower field range (6.40 MV/cm to -4.62 MV/cm) in FeCaps on Sc electrodes, facilities lower leakage and direct observation of FE switching from N-polar to M-polar at positive electric fields on the Sc electrodes as shown in Figure 3b. Due to the high leakage observed for FeCaps made from ultra-thin AlScN films on Al,[37] the ferroelectric switching and resistive leakage peaks overlap at all frequencies, making a direct comparison of the $Ec^+$ between the two devices using AC I-V challenging (Figure S12 a and b). The J-E hysteresis loop without leakage subtraction (Figure S13), however, shows the Sc BE FeCaps have lower $E_C^+$.

To better understand the switching dynamics and improve the reliability of our analysis, we measured $E_C^-$ values vs. frequency on five distinct, 4 μm radius capacitors spatially distributed across the wafer. The mean $E_C^-$ and standard deviation (SD) were calculated, from which symmetric error bars were generated. In Figure 3c, Sc BE FeCaps exhibit lower $E_C^-$ from a 7.1 kHz to 71 kHz range. Notably, a 21.4% $E_C^-$ reduction was observed at 50 kHz. The Kolmogorov–Avrami–Ishibashi (KAI) theory, which has been widely used to describe the switching-frequency dependence of ferroelectrics (Equation 1),[76, 77] was used to describe the dynamic switching response of this ultra-thin AlScN film system. A linear relationship between $\log(|E_C^-|)$ and $\log(f)$ was obtained via logarithmic transformation of Equation 1 into Equation 2. Equation 2 was used to fit the experimental $E_C^-$-frequency data where $E_0$ is the coercive field under DC excitation, $f$ is the experimental frequency, and α is the exponent that characterizes the dependence of $E_C^-$ on frequency. $E_C^-$ is the experimental kinetic negative coercive fields while $E_0^-$ is the negative coercive field under DC excitation obtained from the fitting.

$$E_C = E_0 \cdot f^{\alpha} \tag{1}$$

$$\log(|E_C^-|) = \alpha \cdot \log(f) + \log(|E_0^-|) \tag{2}$$

The logarithmic fitting in Figure 3d shows that, at frequencies below 16.7 kHz, the $E_C^-$ vs. frequency relationship of the Sc BE FeCaps and Al BE FeCaps follows $E_{C^-} \propto f^{0.030}$ and $E_{C^-} \propto f^{0.028}$ respectively (Table S1), suggesting reduced frequency dependence of the coercive field when compared to a previous study reporting measurements of thicker AlScN films grown by Pulsed Laser Deposition (PLD).[78] The measured $|E_0^-|$ values obtained from the fitting were determined to be 2.99 MV/cm for FeCaps with Sc bottom electrode and 3.60 MV/cm for FeCaps with Al bottom electrode (Table S1). Interestingly, at higher frequencies (from 16.7 kHz to 71.4 kHz), the Al BE FeCaps follow $E_{C^-} \propto f^{0.063}$, indicating a stronger frequency dependence of ferroelectric switching for faster external pulses, while Sc BE FeCaps still exhibit an exponent of 0.03, indicating its weaker $E_C$-$f$ dependence at high frequency. This led to an extrapolated $E_C^-$= -4.23 MV/cm for the FeCaps on Sc bottom electrodes at 100 kHz excitation. This $E_C^-$ is significantly lower than the $|E_C^+ + E_C^-|/2$ reported for both $HfN_{0.4}/Al_{0.7}Sc_{0.3}N$(10nm)/TiN and $HfN_{0.4}/Al_{0.7}Sc_{0.3}N$(20nm)/TiN, which were measured to be greater than 8 MV/cm.[48] This shows the benefit of the improved lattice matching of highly Sc alloyed AlScN to the Sc template in reducing $E_C$.

In a relaxor-based ferroelectric material, Chen *et al*. reported a similar kinetic switching change phenomenon to that observed for the AlScN FeCaps on Al bottom electrodes, where the switching dynamics were separated into two frequency regimes due to domain kinetics differences at low and high frequencies.[79] This suggests that the observed behavior in Al BE FeCaps may also be linked to domain kinetics, although further investigation is required to validate this hypothesis. *Yang et al.* experimentally investigated domain wall (DW) motion dynamics under AC fields of $PbZr_{0.2}Ti_{0.8}O_3$ (PZT) systems, where two distinct regimes of $E_C$-frequency dependence were observed across a wide range of operating temperatures. [80] The critical frequency $f_{cr}$, where the $E_C$-$f$ relationship changes, marks the crossover between DW creep motion at low frequencies and viscous flow motion at high frequencies. In the creep regime, polarization switching benefits from sufficient time to overcome energy barriers, whereas in the viscous regime, rapidly increasing external fields provide insufficient time for polarization switching, thereby requiring a higher $E_C$ to cross the energy barrier and causing stronger $E_C$-frequency dependence. In our work, the physical meaning of the $f_{cr}$ for the Al BE FeCaps remains unclear, and the location of the crossover $f_{cr}$ of the Sc BE FeCaps is unknown. However, the weaker $E_C$-frequency dependence and lower $E_C$ of Sc BE FeCaps indicate a reduced switching energy barrier and more efficient switching dynamics. This observation will be the subject of future research examining switching kinetics over broader frequency ranges and as a function of temperature to better resolve the dynamic mechanisms in ultra-thin AlScN systems.

From a theoretical standpoint, *Behrendt et al.* recently used molecular dynamics and Monte Carlo simulations to explain AlN DW migration and growth mechanisms. [60] Their work shows that the rate of nucleation and domain growth plays a major role in the switching process. When the nucleation rate dominates over domain growth, switching behavior follows the conventional KAI model consistent with our observations at low frequencies for both Al and Sc bottom electrodes. However, when nucleation rate is lower relative to the domain growth rate, switching deviates from KAI assumptions and transitions to a process by which domains grow in a fractal manner. Our experimental results align with this theoretical framework. In our work, the deviation between the $E_C$ obtained from DC I-V and the $|E_0^-|$ for AlScN on Al BE FeCaps extracted from the high frequency regime suggests a potential shift in the switching mechanism, which in not observed for the AlScN on Sc BE FeCaps over the studied frequency range. This difference is likely governed by the interplay of nucleation and domain propagation, which could be influenced by strain, interface free energy, applied electric field and defect density.[81-84] Previous Density Functional Theory (DFT) studies in wurtzite systems, such as (Zn,Mg)O, demonstrate how strain fluctuations can modify polarization switching energy barriers. [85] To completely understand the kinetic differences between the AlScN on the two bottom electrodes, more comprehensive simulations incorporating interfacial free energy and defect density contributions should be pursued.

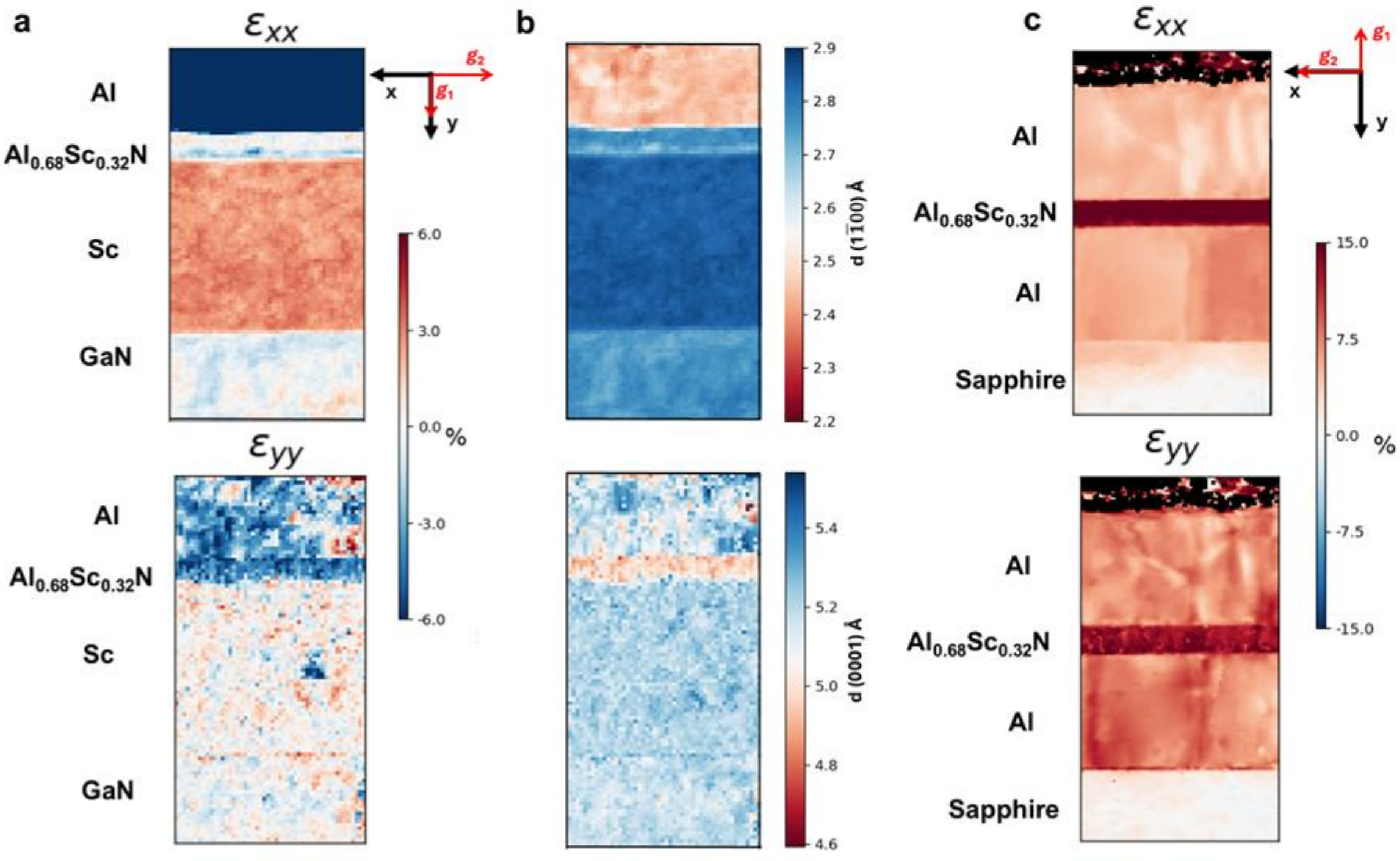

Figure 4. Scanning Electron Nanobeam Diffraction strain mapping of (a) $\boldsymbol{\varepsilon_{xx}}$ and $\boldsymbol{\varepsilon_{yy}}$ of GaN/Sc/$Al_{0.68}Sc_{0.32}N$/Al with respect to GaN, along $[\mathbf{1\bar{1}00}]$ of GaN. (b) *d*-spacing map of GaN/Sc/$Al_{0.68}Sc_{0.32}N$/Al stack along in-plane $[\mathbf{1\bar{1}00}]$ and out-of-plane $[\mathbf{0002}]$ of AlScN, respectively. (c) $\boldsymbol{\varepsilon_{xx}}$ and $\boldsymbol{\varepsilon_{yy}}$ of sapphire/Al/$Al_{0.68}Sc_{0.32}N$/Al with respect to sapphire, along $[\mathbf{1\bar{1}00}]$ of sapphire.

To investigate the effect of the bottom electrode on the structural strain, Scanning Electron Nanobeam Diffraction (SEND) was performed on the entire layer stack of each capacitor.[86] Variations of the reciprocal lattice vector $\vec{g}$ ($|\vec{g}| = 1/d_{spacing}$, where $d_{spacing}$ is the interplanar distance) were measured and then converted to a real space strain tensor,[87] following the mathematical relationship in Equation 3. $\varepsilon_{xx}$ and $\varepsilon_{yy}$ correspond to the in-plane (horizontal) strain tensor and out-of-plane (vertical) strain tensor, calculated using Equation 4 and Equation 5, where $d_{ref}$ is the reference $d_{spacing}$. In this work, the $d_{spacing}$ of GaN and sapphire were selected as the reference $d_{ref}$. These were calculated from the standard lattice parameters from the Joint Committee on Powder Diffraction Files (JCPDF). JCPDF number: 500792 for GaN *a*= 3.18 Å, *c*= 5.18 Å; JCPDF number: 740323 for sapphire *a*=4.76 Å, *c*=12.99 Å. $d_{film}$ is the strained film interplanar spacing.

$$\varepsilon = \frac{d_{film} - d_{ref}}{d_{ref}} = \frac{\Delta d}{d_{ref}} \tag{3}$$

$$\varepsilon_{xx} = \frac{\Delta d_x}{d_{ref_{xx}}} \tag{4}$$

$$\varepsilon_{yy} = \frac{\Delta d_y}{d_{ref_{yy}}} \tag{5}$$

The strain map for the thin films was measured along the along $[\mathbf{1\bar{1}00}]$ and along $[\mathbf{0002}]$ crystallographic directions of AlScN relative to the underlying substrate, shown in Figure 4. For the Sc and Al bottom electrodes, the substrates were GaN and sapphire, respectively. The mapping results show that lattice mismatches between layers create local lattice parameter changes in the electrode films and interfaces, which subsequently influence the lattice parameter of the overlying AlScN films. Notably, in Figure 4a, the AlScN film grown over the Sc bottom electrode exhibits in-plane strain ($\varepsilon_{xx}$ mapping) comparable to GaN due to their better in-plane lattice parameter match. The strain tensor near the interface (bluish region) in the $\varepsilon_{xx}$ map is attributed to the non-uniform interface formed between the Sc bottom electrode and the AlScN film. The $\varepsilon_{yy}$ map shows that AlScN has smaller out-of-plane strain values across the film relative to GaN, which is consistent with the smaller lattice parameter along the [0002] direction. The dark blue at the top

Al layer in $\varepsilon_{xx}$ is due to the cubic symmetry of Al contrasting with the hexagonal closed packed (hcp) structure of the underlying layers, as well as the chosen strain color scale being optimized to better highlight variations in the Sc and AlScN layers. Corresponding lattice parameter maps, shown in Figure 4b, further confirm the better in-plane lattice match between GaN and AlScN, resulting in relatively lower strain between the Sc and AlScN at their interface. In contrast, the Al bottom electrode film exhibits a significant lattice mismatch compared to sapphire (Figure S15), leading to higher $\varepsilon_{xx}$ and $\varepsilon_{yy}$values, as shown in Figure 4c and Figure S16. This larger strain and lattice parameter variation induces greater strain variation across the AlScN grown on the Al bottom electrode in comparison to the more uniform strain distribution observed for the thin film with a Sc bottom electrode. Averaged lattice parameters for each layer of the film stacks were also determined from the SAED patterns and XRD and are summarized in Table 1, XRD values are rounded to match the resolution of the TEM measurements. Additionally, we have also provided the *d*-spacing measurement for other planes in supporting information (Table S2 and Table S3)

**Table I.** Experimentally Measured Lattice Constants Using TEM SAED patterns and XRD

| **Layers** | | **GaN/Sc/AlScN/Al** | | **Sapphire/Al/AlScN/Al** | |
|---|---|---|---|---|---|
| | | TEM | XRD | TEM | XRD |
| Sc | $a$ (Å) | 3.31 | 3.31 | | |
| | $c$ (Å) | 5.28 | 5.28 | | |
| AlScN | $a$ (Å) | 3.24 | - | 3.21 | - |
| | $c$ (Å) | 5.00 | 5.03 | 5.00 | 5.01 |
| Al | $a$ (Å) | 4.03 | 4.05 | 4.03 | 4.04 |

For the strain mapping, we selected GaN and Sapphire substrates as strain mapping references because they are thick, uniform, and defect-free with equilibrium lattice spacings. This choice establishes the measurement scale and does not imply that substrates cause the measured lattice parameter variations in the AlScN. The 4D-STEM results show that lattice mismatch between the layers create local lattice parameter changes in the electrode films, which subsequently influence the lattice parameter of the overlying AlScN films. Although the choice of reference affects the absolute strain quantification due to the $d_{\text{spacing}}$ variations and does not allow a direct distinction between tensile and compressive strain, a clear contrast in the overall residual strain level can still be observed. Specifically, the ultra-thin AlScN film grown on the Sc bottom electrode, which has a smaller lattice mismatch with AlScN, exhibits lower in-plane residual strain compared to the film grown on the Al electrode.

Interestingly, both single-pattern diffraction analysis and strain mapping of the entire layer structure suggest that the in-plane and out-of-plane lattice parameters of ultra-thin $Al_{0.68}Sc_{0.32}N$ vary slightly between the two samples, resulting in a small difference in the *c/a* ratio, which is a key knob to tune the ferroelectricity in AlScN. In the case of AlScN on Sc the *c/a* ratio is ~1.54, while it is ~1.56 for the ultra-thin film grown on Al. Despite this minimal difference in *c/a*, the Sc-based film displays a significantly lower coercive field and a different frequency dependence of $E_C$. Several factors may contribute to this observation. First, reduced interfacial strain may promote improved *c*-axis texture in the AlScN thin film. McMitchell *et al*. reported that modifying strain using different lattice-mismatched growth templates can influence the *c*-axis orientation in ultra-thin AlScN films, thereby affecting both the $E_C$ and the Sc incorporation window in AlN.[38] The XRD measurements, however, do not show a significant difference in the *c*-axis orientation between the two AlScN films. Second, misfit dislocations generated by lattice mismatch and strain in the film may interact with domain walls, leading to a pinning effect that hinders 180° domain wall motion in wurtzite ferroelectric systems, thus suppressing switching behavior.[88-91] The stronger frequency dependence of the $E_C$ at higher frequencies supports higher domain wall pinning in the AlScN films grown on the Al electrode with higher lattice mismatch. The effect of dislocations in PZT films has been widely studied,[92] while this topic is less explored in the AlScN system. To further elucidate the role of dislocations in AlScN, future studies using atomically resolved high-angle annular dark field (HAADF) and annular bright field (ABF) STEM imaging could help identify dislocations with different Burgers vectors near the interface.[93, 94] These potential approaches underscore the importance of defect engineering in the development of scaled AlScN ferroelectric memory device.

CONCLUSION:

Ultra-thin AlScN films were deposited on Sc and Al bottom electrodes with clear *c*-axis orientation based on XRD and electron diffraction patterns. The ferroelectricity of AlScN grown on a Sc bottom electrode was extensively verified and evaluated, showing robust ferroelectricity both under AC and quasi-DC sweeps. By comparing the ferroelectric performance to capacitors with an Al bottom electrode, the Sc BE FeCaps demonstrated $E_C$ reduction 15.8% ($E_C$+) and 10.6% ($E_C$-) under quasi-DC measurements. Furthermore, under AC conditions, the reduction of $E_C$ depends on the frequency. At 50 kHz, the $E_C$- reduction can be more than 20%, indicating lower switching energy barrier realized by smaller-lattice mismatched Sc growth templates.

Applying the KAI model, the dynamic switching behavior of the two capacitors was evaluated. The ferroelectric ultra-thin film grown on Sc maintains a stable frequency dependency of coercive field across a wide frequency range (7.1 kHz to 71 kHz), while the AlScN film grown on Al has a higher frequency dependency for the coercive field at higher frequencies. To understand the mechanism behind the $E_C$ reduction, strain mappings and the in-plane and out-of-plane lattice parameters for the film stacks were measured via SEND for the two samples. These measurements show that the ultra-thin $Al_{0.68}Sc_{0.32}N$ on Sc has lower in-plane strain, a larger *a*-axis lattice parameter and a smaller lattice mismatch with the bottom electrode template. These results suggest that engineering the bottom electrode to minimize lattice mismatch at the interface is a promising strategy to tailor the ferroelectric properties of ultra-thin AlScN films, offering a viable route toward energy-efficient ferroelectric device design.

## EXPERIMENTAL METHODS:

### Film Deposition:

Scandium was selected as a novel growth template for AlScN ultra-thin films and was deposited onto a 5 µm GaN layer on a double side polished (DSP) 4-inch sapphire substrate via physical vapor deposition (PVD) using direct current (DC) sputtering. A 90 nm Sc film was deposited at 500 °C with a 20 standard cubic centimeters per minute (sccm) Ar gas flow under $\sim 1\times10^{-3}$ mbar and a target power of 1.27 W/cm². A 14 nm N-polar $Al_{0.68}Sc_{0.32}N$ film was then synthesized in the same chamber without a vacuum break to prevent the oxidation of the bottom metal film. The AlScN ultra-thin film was deposited in a 30 sccm $N_2$ atmosphere (pressure range from $8\sim8.3\times10^{-4}$mbar) with an Al target power of 12.7 W/cm$^2$ and a Sc target power 8.32 W/cm$^2$ via reactive pulsed DC co-sputtering. This deposition was followed by the creation of a top Al electrode at 150 °C under a 20 sccm Ar atmosphere with a target power of 12.7 W/cm$^2$.

For comparing of the ferroelectric performance of ultra-thin $Al_{0.68}Sc_{0.32}N$ films, a 50 nm Al bottom electrode was deposited on sapphire under the same conditions described above. An ∼12 nm thick $Al_{0.68}Sc_{0.32}N$ was subsequently synthesized followed by a 50 nm Al top electrode deposition. The entire film stack was deposited without breaking vacuum to prevent oxidation.

Device Fabrication

To fabricate the capacitors, photoresist on the Al top electrode was patterned with photolithography to construct 4 μm radius devices. This was followed by $BCl_3/Cl_2$ plasma etching, using an Oxford PlasmaPro 100 Cobra inductively coupled plasma (ICP) etcher, which also resulted in the exposure of the bottom electrode. An ∼90 nm thick layer of $SiO_2$ was selected as a dielectric material to create a larger probing pad, thereby avoiding direct contact of the probe with the thin ferroelectric capacitors. $SiO_2$ was deposited by chemical vapor deposition (CVD) at 200 ºC. Next, vias to each capacitor were constructed by photolithography and $CF_4$ plasma dry etching of the $SiO_2$. An Ar plasma sputter etch was used to remove the alumina from the Al top electrode surface and was followed by in-situ ∼100 nm Al deposition. The pad metal was then patterned by lithography and a $BCl_3/Cl_2$ plasma etching process. Device dimensions were acquired by Zeiss Axio Imager M2m Microscope.

Electrical Measurements:

Electrical measurements were conducted using a Keithley 4200A-SCS analyzer. For Alternative current (AC) measurement, triangular pulse J-E hysteresis loops with different frequencies were conducted on the FeCaps to verify the ferroelectricity. Triangular Positive-Up-Negative-Down (PUND) tests were performed on the devices with two types of bottom electrode to identify the $E_C$ for comparison analysis. 500 ns ultra-fast rectangular PUND pulses were conducted on the Sc BE FeCaps to study the polarization switching. For DC I-V hysteresis analysis, a quasi-DC I-V (~0.01 Hz) voltage sweep with a 0.025 V step size was applied on the two FeCaps to analyze the $E_C$ under quasi-DC conditions. All the tests were conducted with a 0.2-micron diameter, 2" long tungsten flexible cat-whisker probe tip for the top electrode and a standard 0.6 micro diameter tungsten probe tip for the bottom electrode.

Material Characterization:

For the cross-sectional TEM imaging of the thin films, a thin lamella suitable for the TEM analysis was prepared using a dual beam focused ion beam microscope (FEI Strata DB235). The lamella was thinned down to ~1 µm using a Ga ion beam at 30 kV and 1 nA, followed by thinning down to ~100 nm at 30 kV and 100 pA current. To remove the damaged and amorphous surface layer, a low kV cleaning of the lamella was performed using 5 kV and 30 pA ion beam. TEM imaging and scanning electron nanobeam diffraction (SEND) analysis were performed using a JEOL F200 microscope operating at 200 kV equipped with two image-forming cameras: OneView and low-dose electron counting Metro camera provided by GATAN. TEM images and SAED patterns were recorded using the OneView camera. A convergence angle of 2 mrad and 100 mm camera length were used to record the diffraction patterns (256 X 256) using the Metro camera. The data were further processed using py4DSTEM v. 0.14.18 provided by the National Center for Electron Microscopy, Lawrence Berkeley National Laboratory.[95]

X-ray diffraction (XRD) and reflectivity (XRR) measurements were performed using a Panalytical X'Pert$^3$ MRD diffractometer equipped with a Ge (220) double-bounce monochromator, a PIXcel detector, and a triple-axis point detector. Lattice parameters were extracted from peak fitting of XRD reflections, while XRR data were analyzed using GSAS-II, which employs the optical matrix method.[96] The triple-axis point detector provides higher angular resolution, but at the expense of intensity. This reduced signal during peak fitting leads to larger error bars, derived from 95% confidence interval. For example, $\theta$-$2\theta$ scans using the triple-axis point detector yielded *c*-lattice parameter uncertainties of +/- 0.021 Å (AlScN on Sc) and +/- 0.013 Å (AlScN on Al). By contrast, measurements of AlScN on Al using the PIXcel detector, which achieve much higher counts, reduced the error to +/- 0.001 Å – an order of magnitude improvement. Rocking curve measurements were also conducted to determine the texture quality of the Sc and AlScN films by fitting the rocking curve of the 0002 peak.

Corresponding Author

stach@seas.upenn.edu, rolsson@seas.upenn.edu

Author Contributions

Y.Z., R.R., E.S. and R.O. conceived the study. Y.Z. optimized film growth conditions, synthesized the metal-capacitor-metal film stacks, fabricated the devices, and measured the ferroelectric properties and R.O. supervised her. R.R performed the TEM characterization and strain mapping and E.S. supervised him. G.E. conducted XRD and XRR experiments. Y.W. assisted with fitting and methodology of electrical measurements under the supervision of D.J. Y.Z., R.R. and G.E. wrote the manuscript. All authors assisted with manuscript revisions. All authors have given approval to the final version of the manuscript.

Funding Sources

This work was partially supported by Intel Corporation under the SRS program. This work is supported in part by Army/ARL via the Collaborative for Hierarchical Agile and Responsive Materials (CHARM) under cooperative agreement W911NF-19-2-0119. This work was carried out in part at the Singh Center for Nanotechnology, which is supported by the NSF National Nanotechnology Coordinated Infrastructure Program under Grant NNCI-2025608. Additional support to the Nanoscale Characterization facility by the NSF through the University of Pennsylvania Materials Research Science and Engineering Center (MRSEC) (DMR-2309043) is acknowledged. This article has been authored by an employee of National Technology & Engineering Solutions of Sandia, LLC under Contract No. DE-NA0003525 with the U.S. Department of Energy (DOE). The employee owns all right, title and interest in and to the article and is solely responsible for its contents. The United States Government retains and the publisher, by accepting the article for publication, acknowledges that the United States Government retains a non-exclusive, paid-up, irrevocable, world-wide license to publish or reproduce the published form of this article or allow others to do so, for United States Government purposes. The DOE will provide public access to these results of federally sponsored research in accordance with the DOE Public Access Plan https://www.energy.gov/downloads/doe-public-access-plan.

# (Supporting Information)

# Coercive Field Reduction in Ultra-thin $Al_{1-X}Sc_XN$ via Interfacial Engineering with a Scandium Electrode

*AUTHOR NAMES Yinuo Zhang,* [1] *Rajeev Kumar Rai,* [2] *Giovanni Esteves,* [3] *Yubo Wang,*[1] *Deep M. Jariwala,*[1] *Eric A. Stach* [2,]* *and Roy H. Olsson, III* [1,]*

AUTHOR ADDRESS

1. Department of Electrical and Systems Engineering, University of Pennsylvania, Philadelphia, PA, USA 19104

2. Department of Materials Science and Engineering, University of Pennsylvania, Philadelphia, PA, USA 19104

3. Microsystems Engineering, Science and Applications (MESA) Center, Sandia National Laboratories, Albuquerque, NM, USA 87185

* Corresponding Authors: stach@seas.upenn.edu, rolsson@seas.upenn.edu

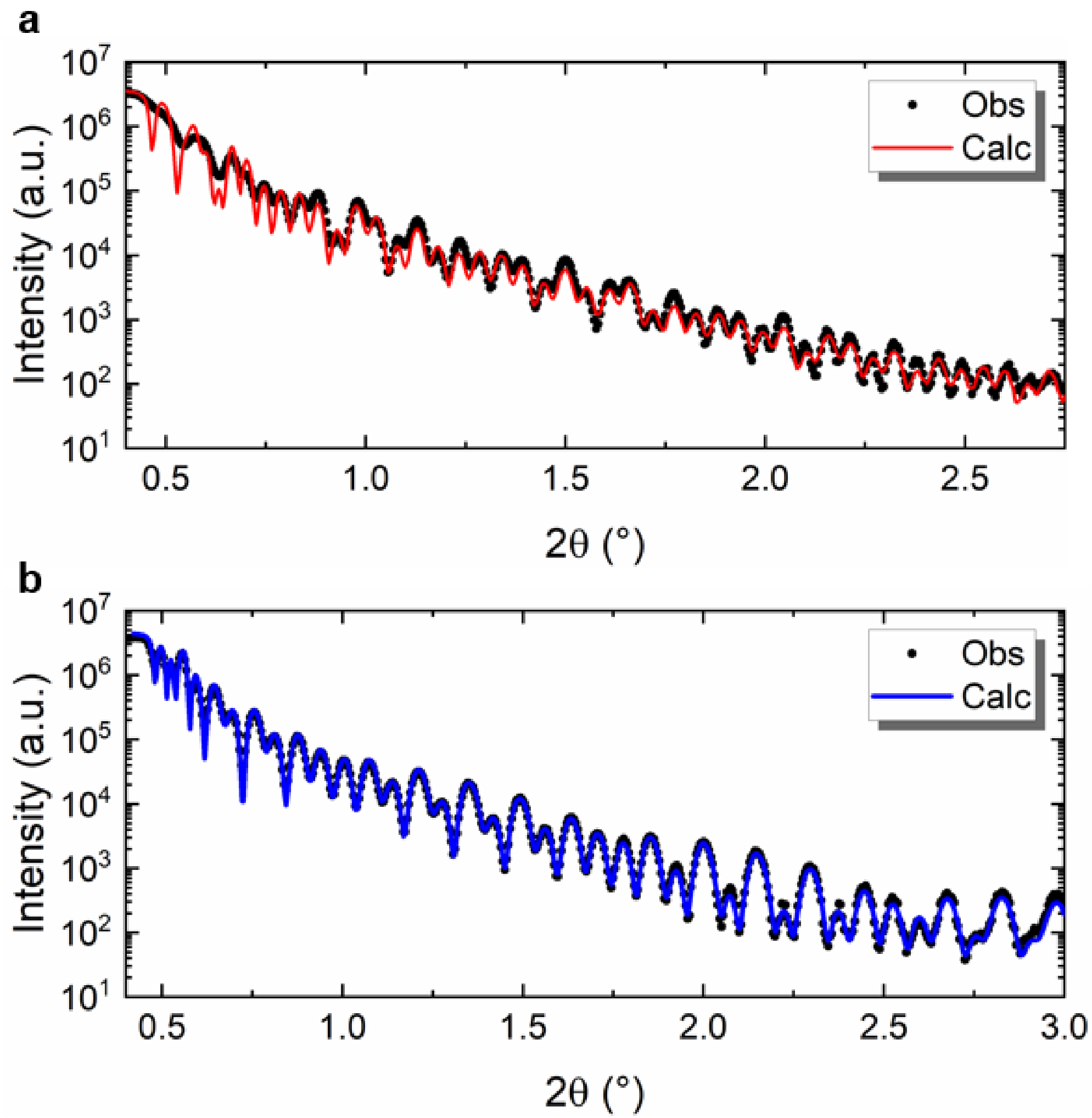

**Figure S1.** (a) XRR spectrum with fitting line (red) of AlScN on the Sc bottom electrode, obtaining a thickness 14.68 nm for AlScN (b) XRR spectrum with fitting line (blue) of AlScN on the Al bottom electrode, obtaining 10.77 nm thickness for the AlScN.

The fit for the Sc/AlScN electrode required modeling three distinct Al layers due to surface defects that caused increased roughness. Including these additional Al layers improved the fit of the XRR at angles below 1°. In contrast, the Al/AlScN sample did not exhibit such defects and was refined using the expected Al/AlScN/Al three-layer model. The error bars of ±1 nm reflect variations from multiple least-squares refinement approaches applied to the XRR data.

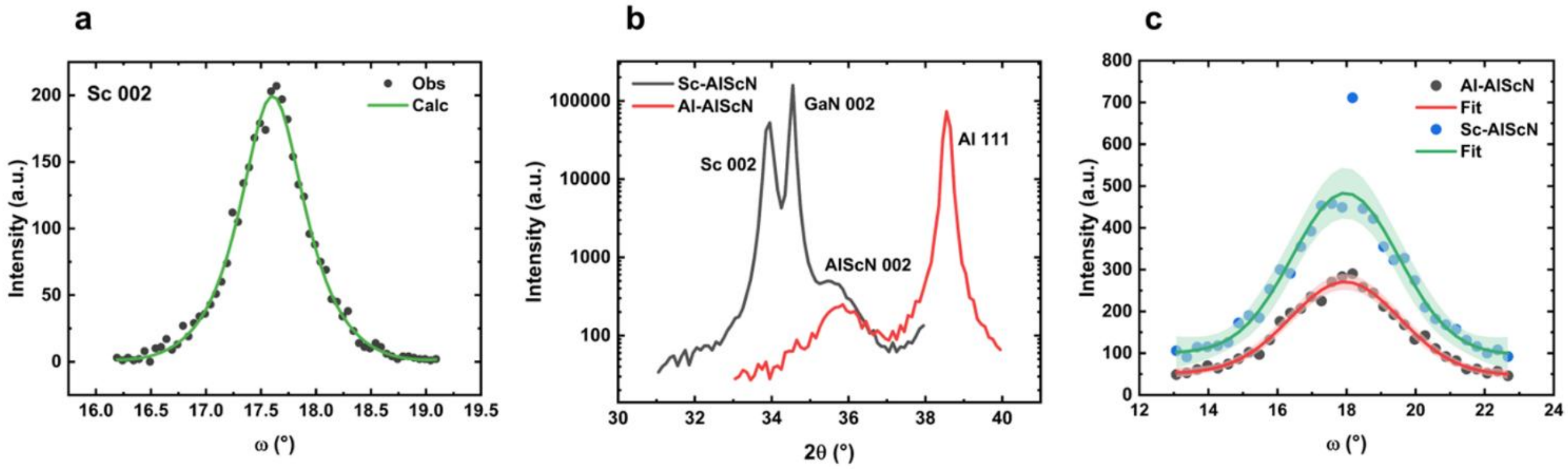

**Figure S2.** (a) Rocking curve scan of Sc 002, obtained a FWHM of 0.76° from peak fitting. (b) $\boldsymbol{\theta/2\theta}$ scan from 30 ° to 40 ° of GaN/Sc/AlScN/Al and Al/AlScN/Al films by using a triple-axis point detector, showing Sc 002, GaN 002, AlScN 002 and Al 111 peaks. (c) Rocking curve scan of 002 AlScN on Sc and Al bottom electrode, generating full-width-half-maximum (FWHM) 3.62°±0.4 for Sc/AlScN/Al and 3.73°±0.23 for Al/AlScN/Al. A Gaussian function was employed to fit both rocking curve scans. The high-intensity point observed in the Sc-AlScN sample is attributed to the GaN substrate, and the use of a point detector significantly reduces its intensity. Notably, this high-intensity point does not appear to skew the fit upwards. The resulting fits are plotted in red and green, with error bars representing the 95% confidence interval for the fits.

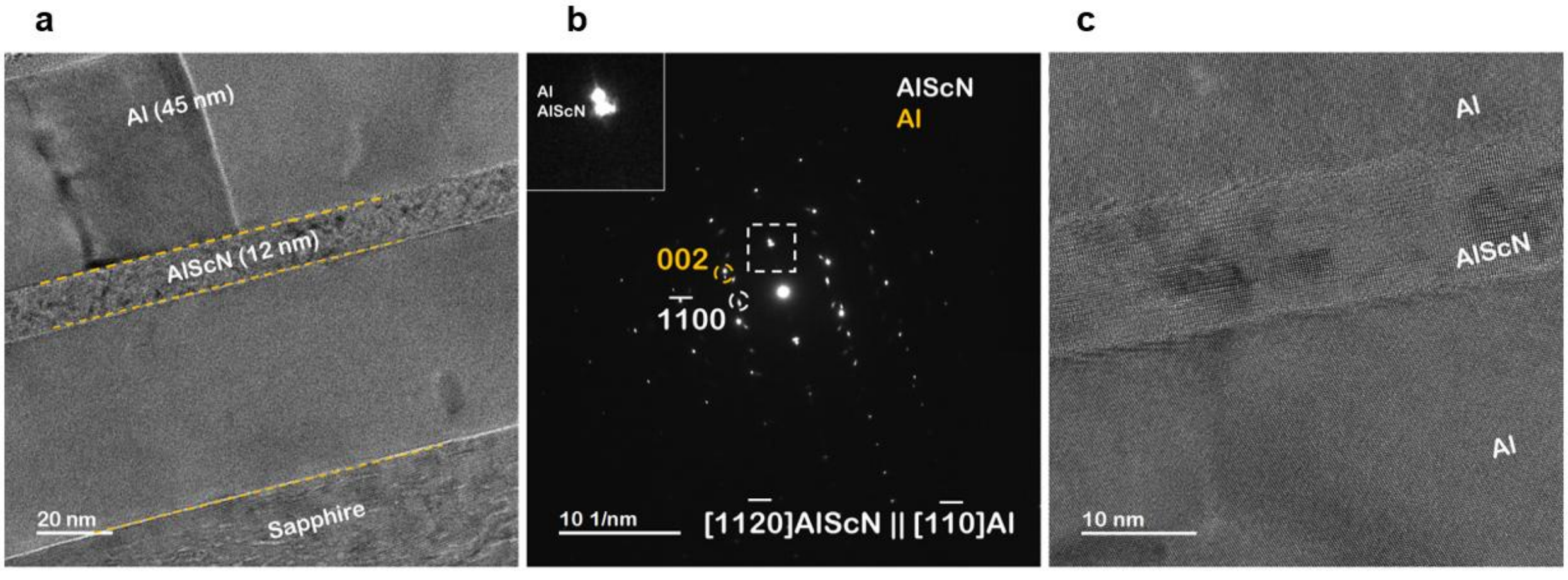

**Figure S3.** (a) Cross-sectional low–magnification TEM image showing the various thin film layers of sapphire/Al/AlScN/Al. The interfaces between various layer have been marked by yellow dotted lines (b) SAED pattern from the Al/AlScN/Al region along [11$\bar{2}$0] of AlScN and [1$\bar{1}$0] of Al, inset: zoomed in out-of-plane spots showing the 2 different lattice spacing corresponding to (0002) of AlScN and (111) of Al planes. (c) High-magnification image of AlScN and interfaces between Al & AlScN.

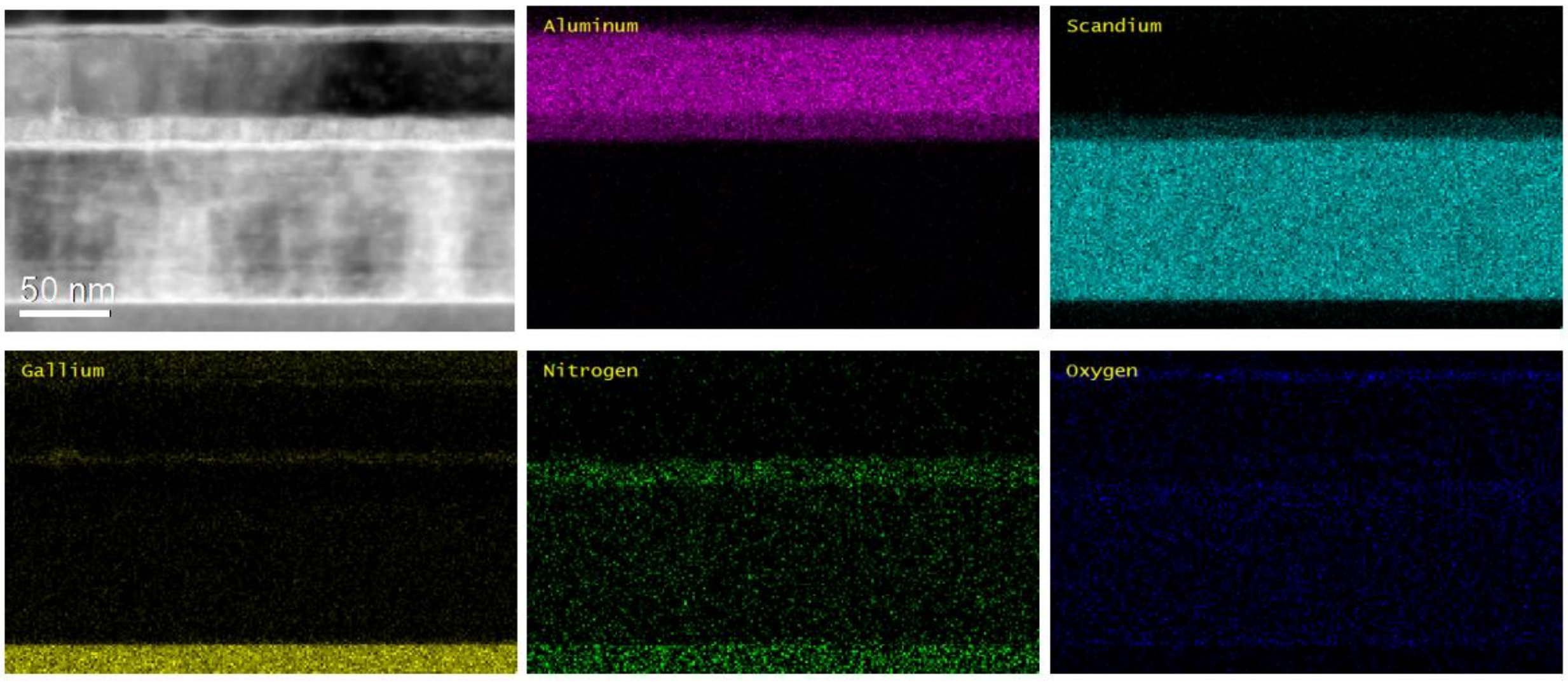

**Figure S4.** EDS mapping of the TEM cross-section of GaN/Sc/$Al_{0.68}Sc_{0.32}N$/Al, showing elemental distributions of aluminum, scandium, gallium, nitrogen and oxygen.

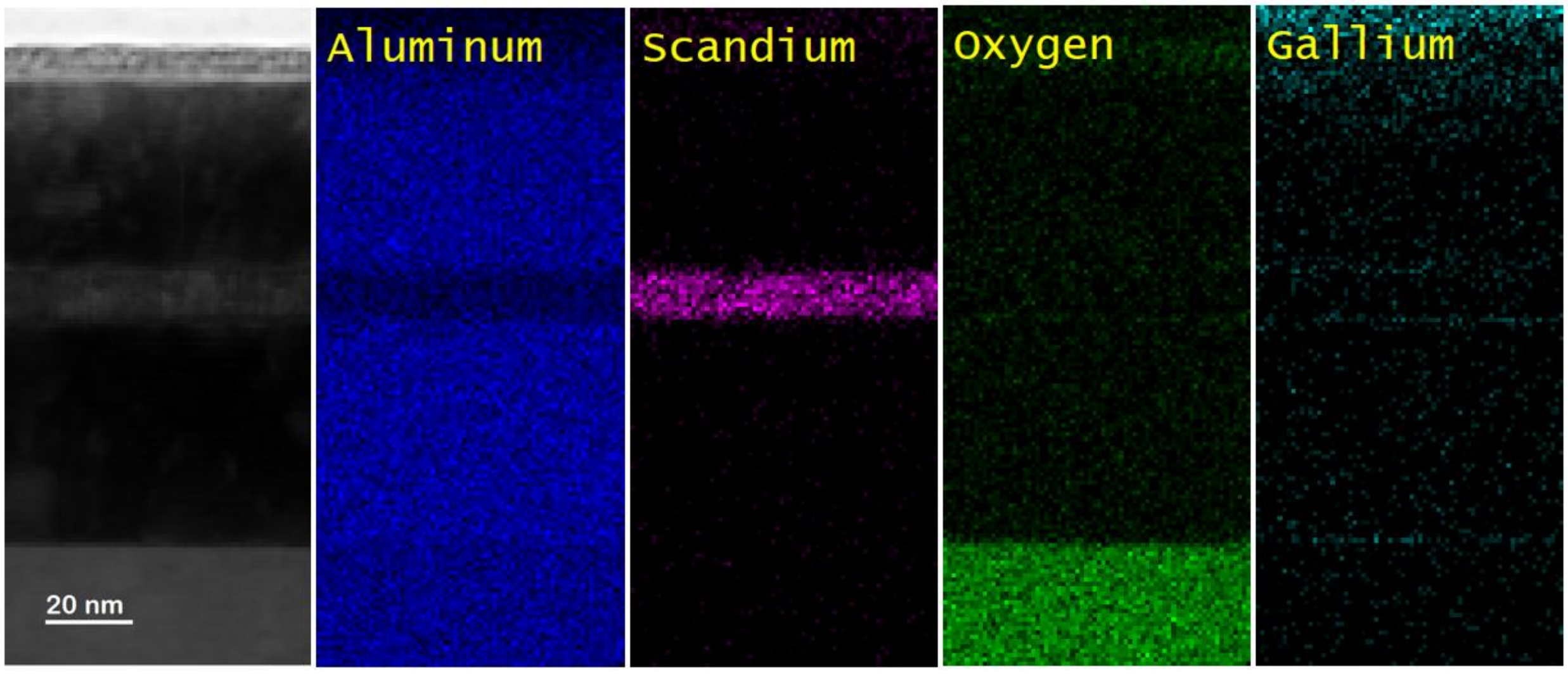

**Figure S5.** EDS mapping of the TEM cross-section of Sapphire/Al/$Al_{0.68}Sc_{0.32}N$/Al, showing elemental distributions of aluminum, scandium, gallium, nitrogen and oxygen.

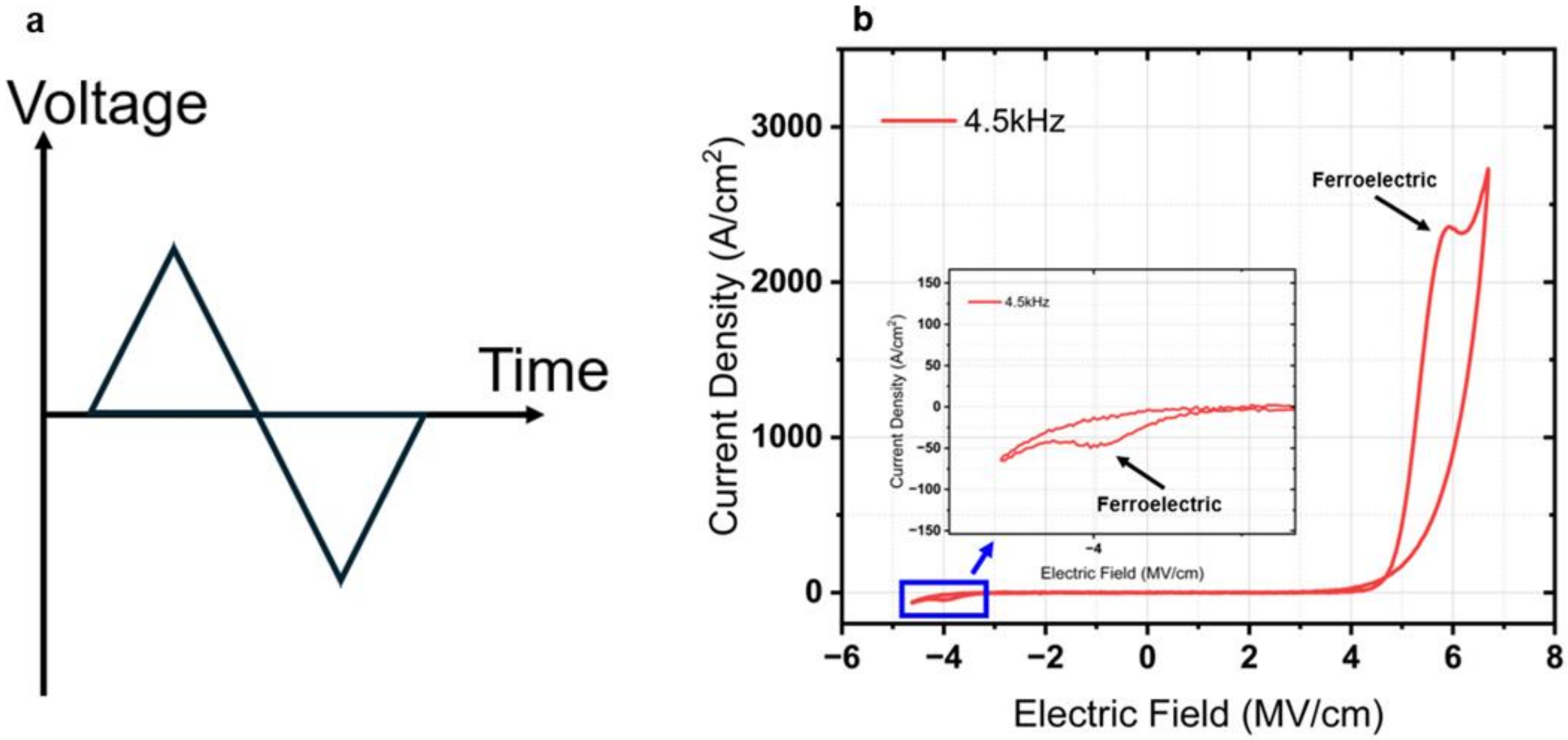

**Figure S6.** (a) Triangular waveform J-E loop hysteresis loops. (b) Current density vs. electric field of Sc/$Al_{0.68}Sc_{0.32}N$ /Al ferroelectric capacitors under 4.5 kHz, with black arrows pointing to ferroelectric current due to dipole switching. Zoomed-in is the current density behavior at negative applied electric field.

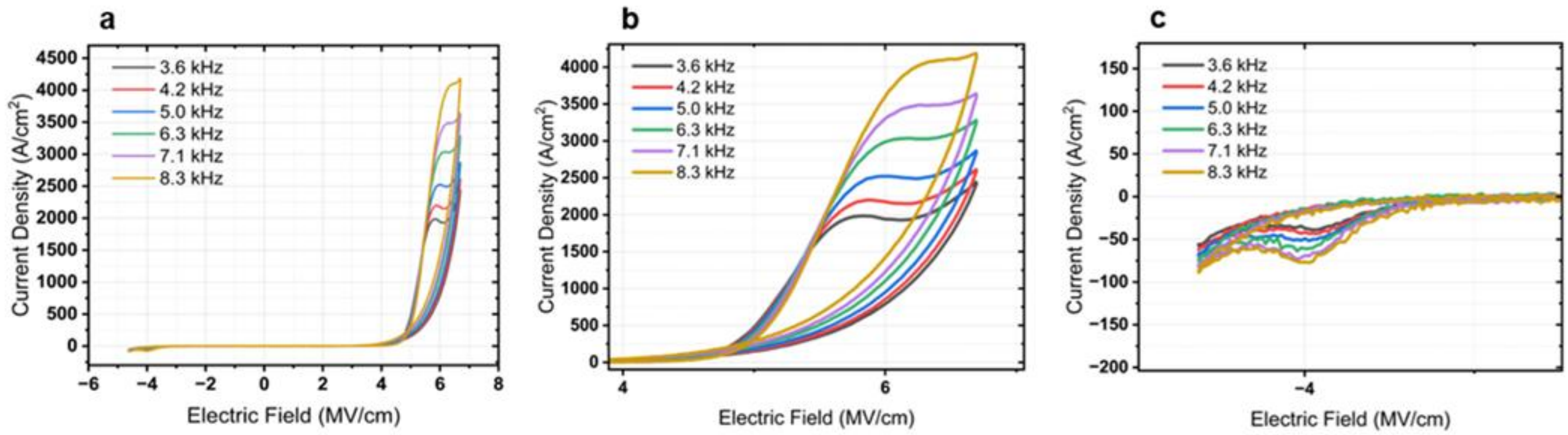

**Figure S7.** Ferroelectricity vs. frequency under AC I-V test. (a) J-E hysteresis loop of Sc/$Al_{0.68}Sc_{0.32}N$ /Al ferroelectric capacitors under different frequencies. (b) Zoomed-in ferroelectric switching peaks and resistive leakage at positive applied electric fields. (c) Zoomed-in ferroelectric switching and resistive leakage at negative electric fields.

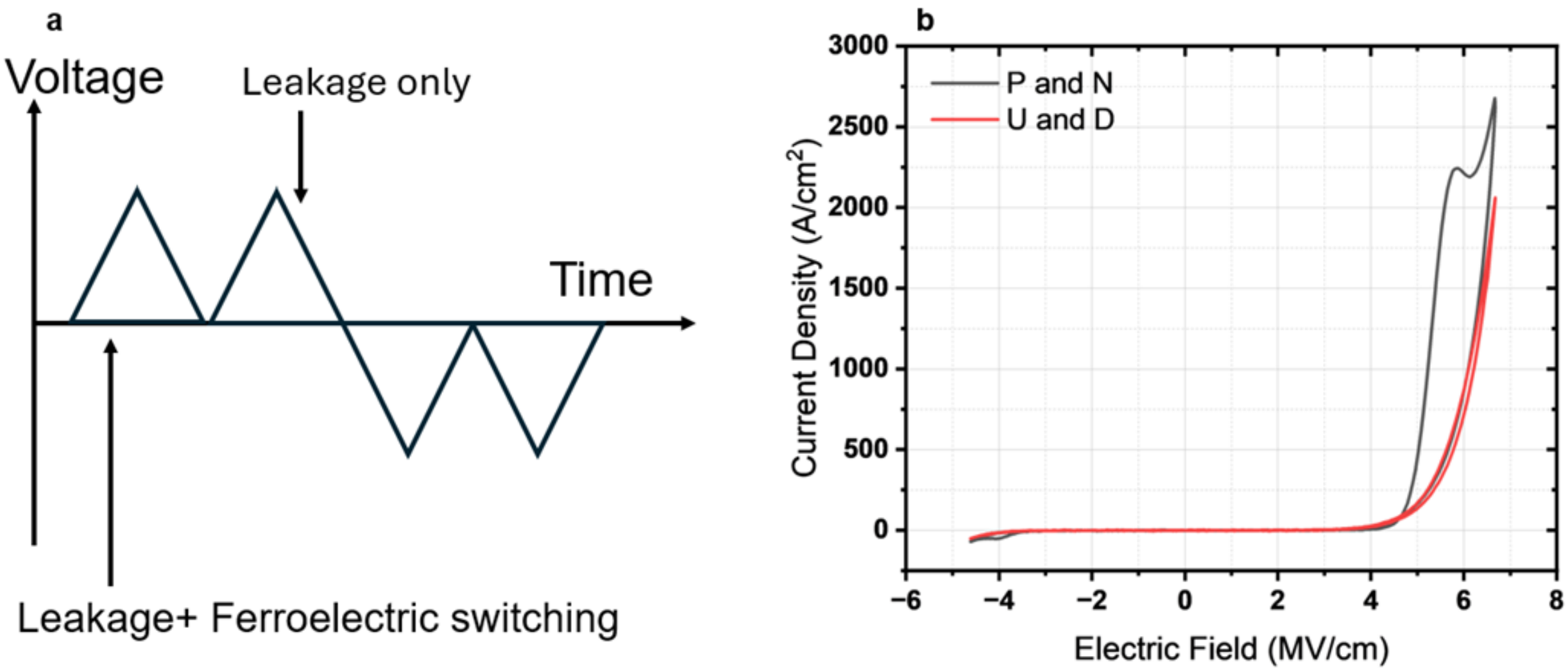

**Figure S8.** (a) Waveform of triangular PUND J-E test. (b) Triangular PUND J-E hysteresis loops under 4.5 kHz for $Sc/Al_{0.68}Sc_{0.32}N/Al$ capacitors before leakage subtraction.

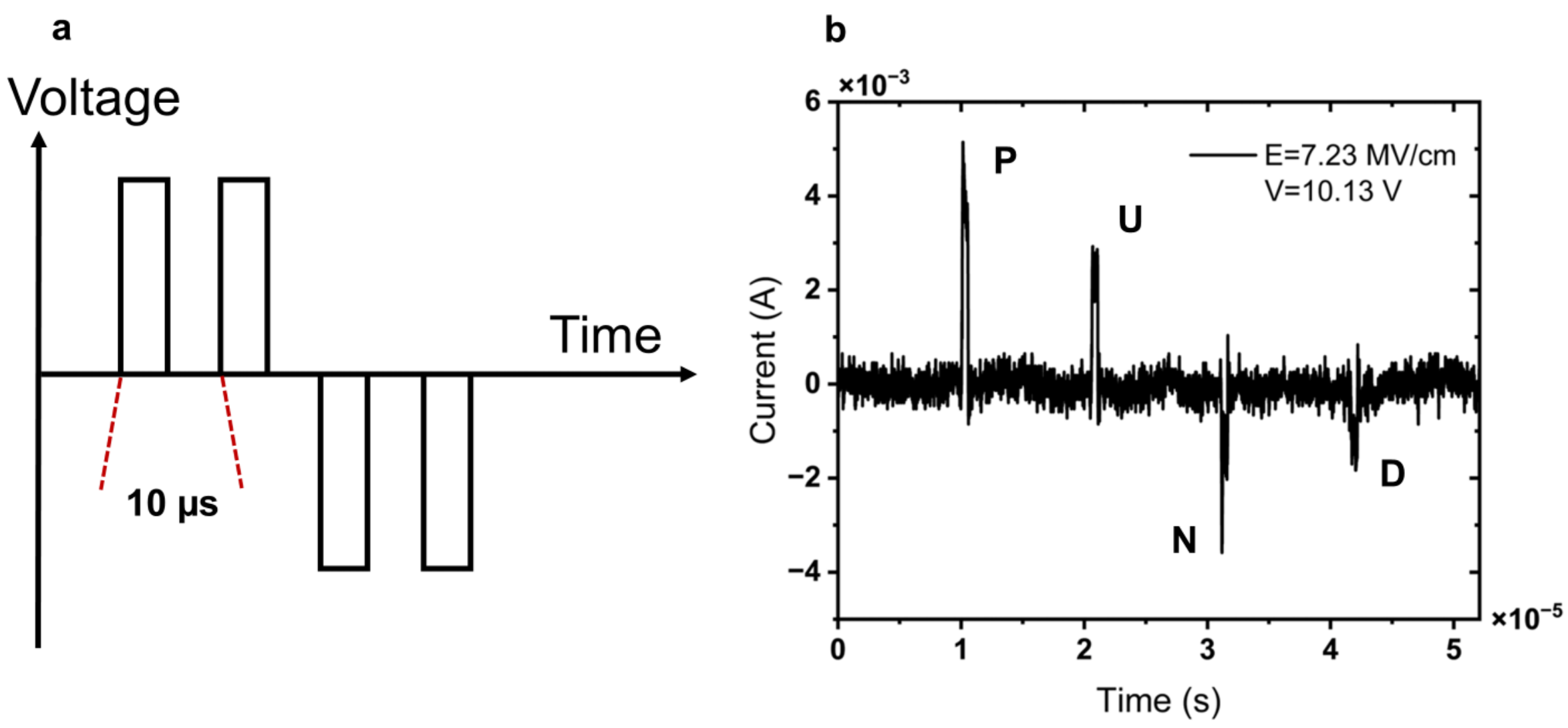

**Figure S9.** (a) Rectangular waveform of ultra-fast 500 ns PUND measurement. (b) Current-time response of a $Sc/Al_{0.68}Sc_{0.32}N/Al$ capacitor from PUND measurement with an applied voltage of 10.13 V.

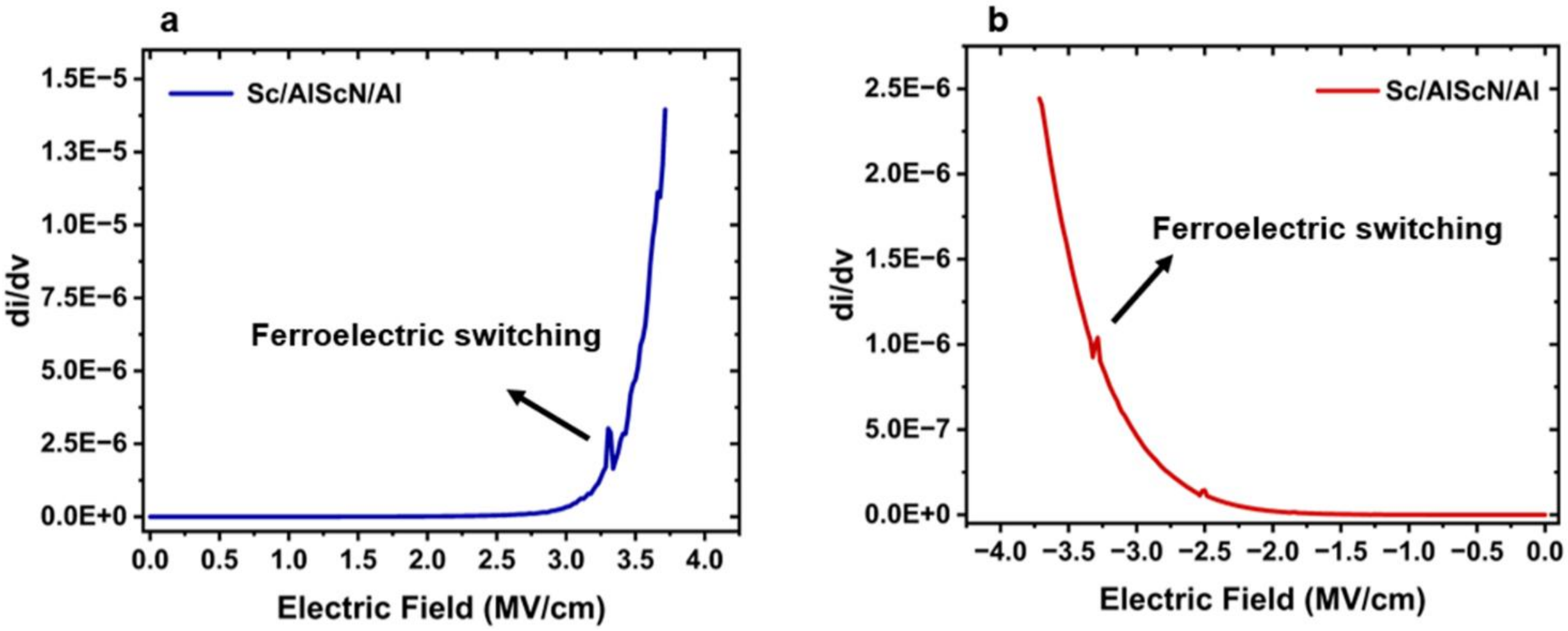

**Figure S10.** $E_C$ identification from a quasi- DC I-V sweep of a Sc/$Al_{0.68}Sc_{0.32}N$/Al capacitor. (a) First derivative of current with respect to voltage vs. applied electric field for path 1 in Fig. 2i. (b) First derivative of current with respect to voltage vs. applied electric field for path 3 in Fig. 2i.

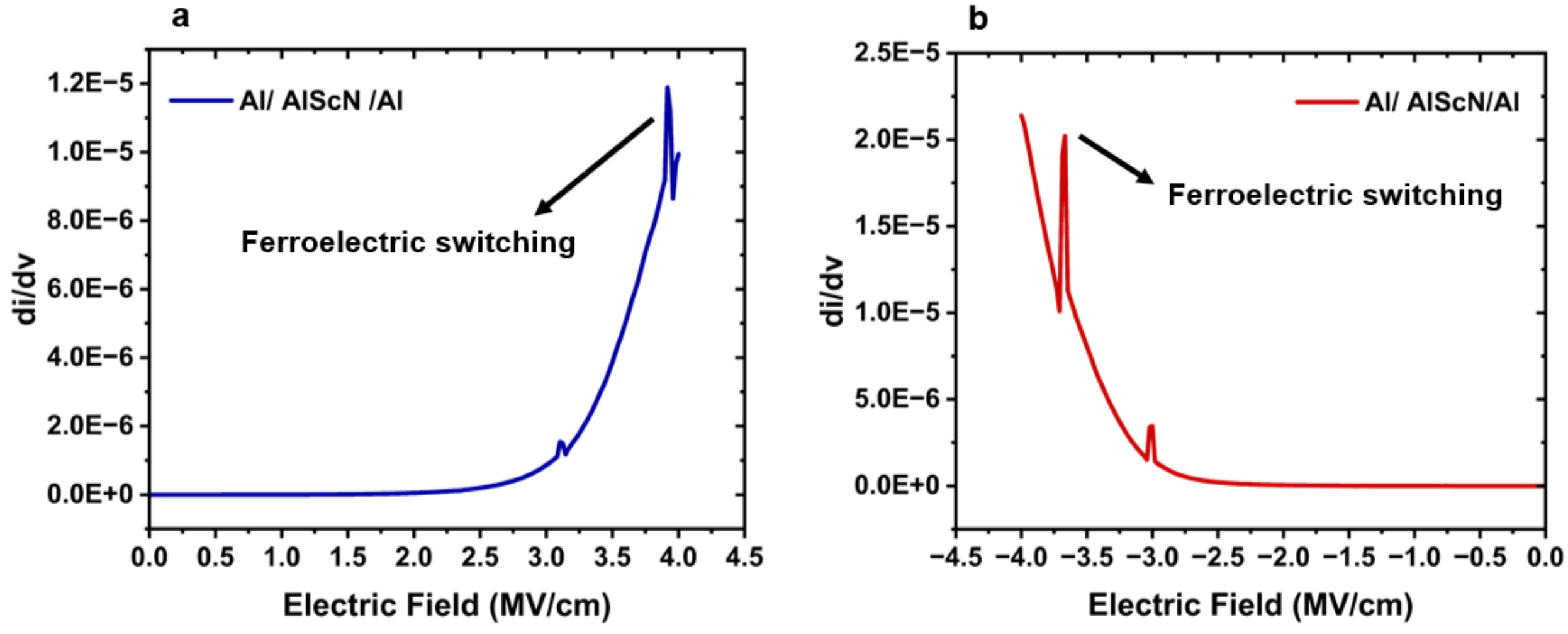

**Figure S11.** $E_C$ location from a quasi- DC I-V sweep of an Al/$Al_{0.68}Sc_{0.32}N$/Al capacitor. (a) First derivative of current with respect to voltage vs. applied electric field for path 1 in Fig. 3a. (b) First derivative of current with respect to voltage vs. applied electric field for path 3 in Fig. 3a.

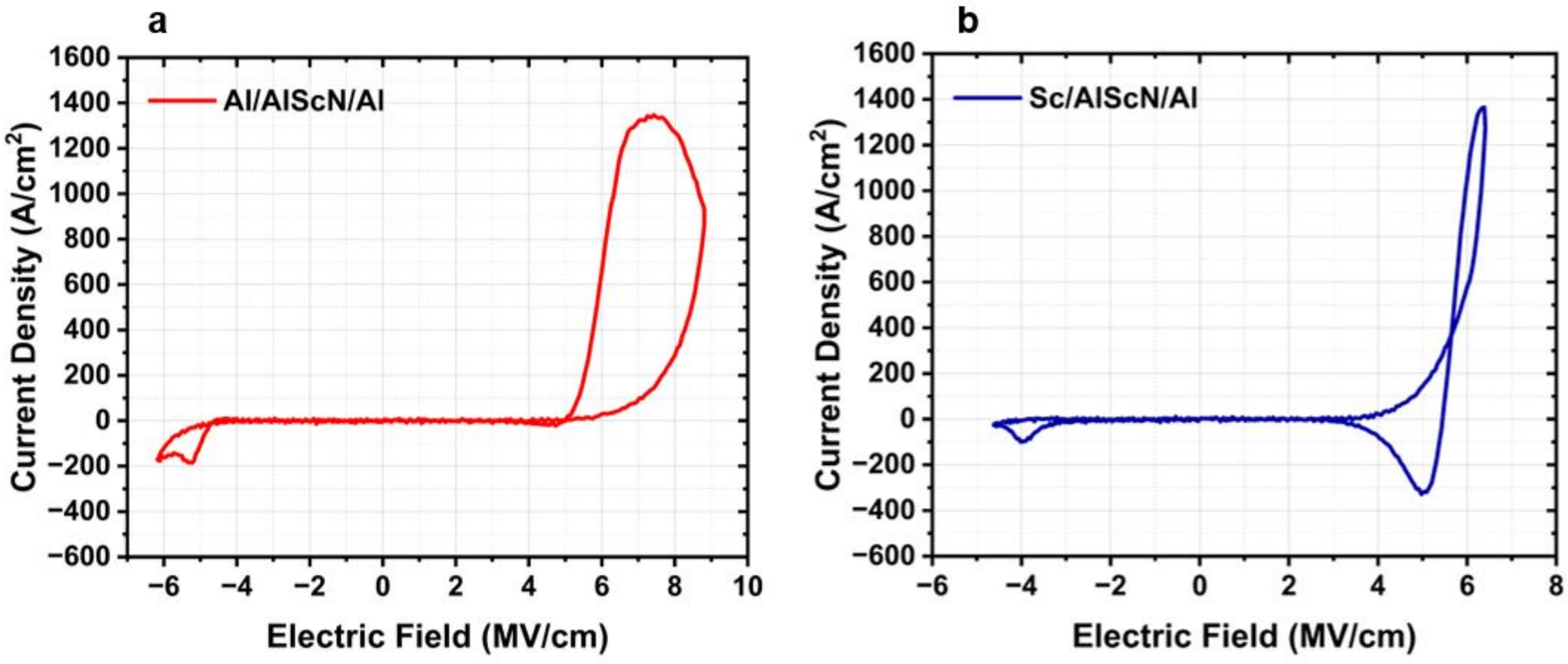

**Figure S12.** 25 kHz Triangular PUND J-E hysteresis loop with leakage compensation. (a) 8.80 MV/cm to -6.17 MV/cm sweeping loop of an Al/$Al_{0.68}Sc_{0.32}N$/Al capacitor. (b) 6.40 MV/cm to -4.62 MV/cm sweeping loop of a Sc/$Al_{0.68}Sc_{0.32}N$/Al capacitor.

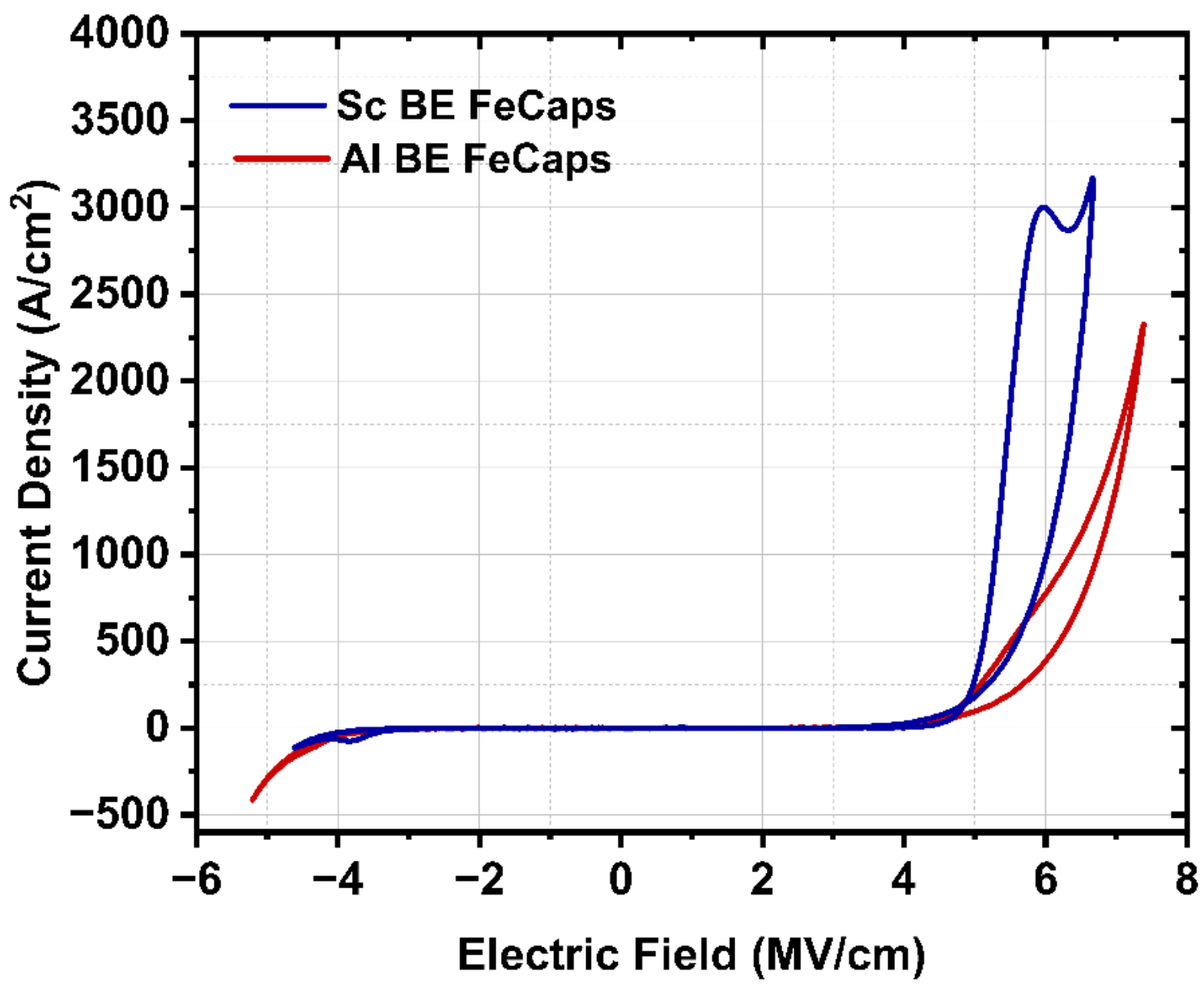

**Figure S13.** 7.1 kHz Triangular PUND J-E hysteresis loop without leakage compensation of Sc BE FeCaps and Al BE FeCaps, showing the $E_C^+$ difference between the two devices.

***Supplementary Note 1. Discussion of Asymmetric Currents.***

Asymmetry in displacement current is commonly observed in sputtered AlScN thin films. For example, Schönweger *et al*. reported peak current densities of approximately 1000 $A/cm^2$ and -2000 $A/cm^2$ post leakage compensation for a 5 nm thick AlScN film on a Pt electrode excited by a 100 kHz triangle wave.[1] Ryoo *et al*. studied the thickness scaling of AlScN on $HfN_{0.4}$ electrodes where small but varying levels of current asymmetry were observed.[2] For example, for a 20 nm thick film excited with a 100 kHz triangle wave, the peak currents were approximately 1350 $A/cm^2$ and -1100 $A/cm^2$. In our work, the peak current densities after leakage compensation using a similar 71 kHz frequency triangle wave for the 14 nm thick AlScN on Sc FeCaps are 1795 $A/cm^2$ and -230 $A/cm^2$ and for the 12 nm thick AlScN on Al FeCaps are 1943 $A/cm^2$ and -409 $A/cm^2$. These can be seen in Figure S14. Notably, while the peak displacement currents under positive electric fields are similar to prior works, the displacement currents under negative electric fields are much smaller and sharper, especially on the Sc electrodes. This can be attributed to the very low leakage currents observed when applying negative electric fields, where leakage compensation minimally alters the JE hysteresis loop as seen in Figure S14(f).

One underlying cause of the current asymmetry is the somewhat higher coercive field for N-polar to M-polar switching. This can be estimated to be 5.75 MV/cm for positive fields from the crossover point observed in the positive field JE hysteresis loop on Sc electrodes highlighted in Figure S14(e). This is compared to -4.22 MV/cm identified by the peak in the JE loop under negative electric fields on the Sc electrodes in Figure S14(f). Due to this, a slightly higher electric field was intentionally applied for positive voltages to ensure full ferroelectric switching in the presence of the imprint in the as deposited films. The lower electric field required to ferroelectrically switch the M-polar to N-polar transition results in exceptionally low leakage here, and it is this low leakage that is primarily the source of the asymmetry compared to prior works.

*Guido et al.* reported that by running cycling on $Al_{1-X}Sc_XN$ ferroelectric capacitors, the imprint can be reduced and asymmetric ferroelectric switching currents can become more balanced. This effect is attributed to defect generation and re-distribution during cycling, promoting a more energetically preferable and faster N-to-Metal-polar switching. Future studies will investigate the imprint behavior as well as ferroelectric current asymmetry mitigation by performing field cycling in this ultra-thin $Al_{1-X}Sc_XN$.[3]

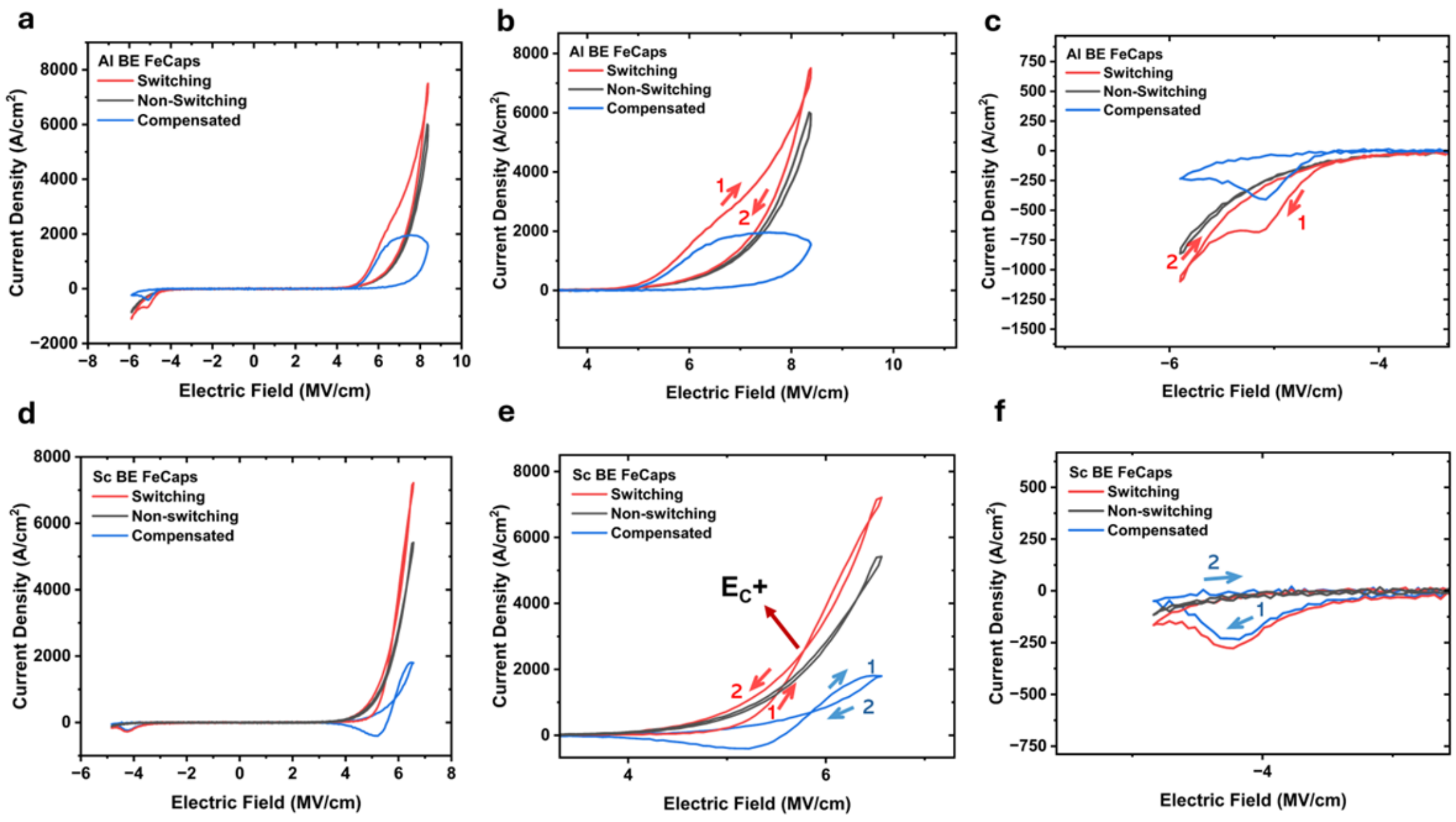

**Figure S14.** PUND J-E hysteresis behavior at 71 kHz before and after leakage subtraction of FeCaps with two types bottom electrodes. (a) PUND J-E sweeps with and without leakage compensation of Al BE FeCaps. (b) Positive loops with and without leakage subtraction of Al BE FeCaps where arrows indicate the sweep direction. (c) Negative PUND J-E sweeps with and without leakage subtraction of Al BE FeCaps, where arrows indicate the sweep direction. (d) PUND J-E sweeps with and without leakage compensation of Sc BE FeCaps. (e) Positive loops with and without leakage subtraction of Sc BE FeCaps where arrows indicate the sweep direction. (f) Negative PUND J-E sweeps with and without leakage subtraction of Sc BE FeCaps, where arrows indicate the sweep direction after leakage subtraction.

An additional item that requires addressing in the 71 kHz data is the negative leakage compensated currents at positive electric fields on the Sc electrodes in Figure S14(d-e). Due to the ferrodiode effect, first reported by Liu *et al*. for AlScN, the leakage current when the applied electric field is aligned with the ferroelectric polarization is much higher than when the applied field and polarization are misaligned.[4] Thus, on the switching sweep as the electric field is increased from 0 towards $E_C$, the polarization and applied E-field are misaligned and the leakage current is lower for the same applied electric field when compared to the non-switching sweep and the portion of the switching sweep when the field is swept from $E_C$ to 0. In both these cases, since the material has been fully switched M-polar, the applied field and polarization are aligned and the leakage current is higher. These different high and low restive leakage states are well described by a Poole–Frenkel tunneling model.[5] Thus, when the high leakage current below $E_C$ for the non-switching measurement is subtracted from the low leakage current below $E_C$ for the switching measurement, an anomalous negative compensated current is obtained. Similarly, the ferrodiode effect confounds extraction of the positive $E_C$ from the leakage compensated measurements of AlScN on Al electrodes.

Yet another phenomenon impacting the leakage compensation in these measurements is the higher leakage current just after ferroelectric switching during the switching waveform when compared to the non-switching waveform. This has been commonly observed in our prior studies.[6-8] The source of this difference in leakage is a subject of on-going research.

**Table S1.** Lograthmic fitting of $|E_{C-}|$ and frequency information of FeCaps with different bottom electrodes

| Bottom Electrode | $Log(\|E_0^-\|)$ | Standard Deviation ($Log(\|E_0^-\|)$) | $\|E_0^-\|$ (MV/cm) | α | Standard Deviation(α) | $R^2$ |
|---|---|---|---|---|---|---|
| Sc | 0.476 | ± 0.019 | 2.99 | 0.030 | ±0.0046 | 0.85 |
| Al (low frequency) | 0.565 | ± 0.010 | 3.60 | 0.028 | ±0.0024 | 0.98 |
| Al (high frequency) | 0.409 | ± 0.017 | 2.56 | 0.063 | ±0.038 | 0.98 |

**Table S2.** Measured *d*-spacing for other planes in GaN/Sc/$Al_{0.68}Sc_{0.32}N$/Al thin film stack

| Planes | d-spacing (Å) | |
|---|---|---|
| | Sc | AlScN |
| 100 | 2.86 | 2.79 |
| 102 | 1.94 | 1.87 |
| 103 | 1.50 | 1.43 |

**Table S3.** Measured *d*-spacing for other planes in Sapphire/Al/$Al_{0.68}Sc_{0.32}N$/Al thin film stack

| AlScN planes | d-spacing (Å) | Al Planes | d-spacing (Å) |
|---|---|---|---|
| 100 | 2.77 | 111 | 2.33 |
| 101 | 2.44 | 200 | 2.03 |
| 102 | 1.86 | | |
| 103 | 1.43 | | |

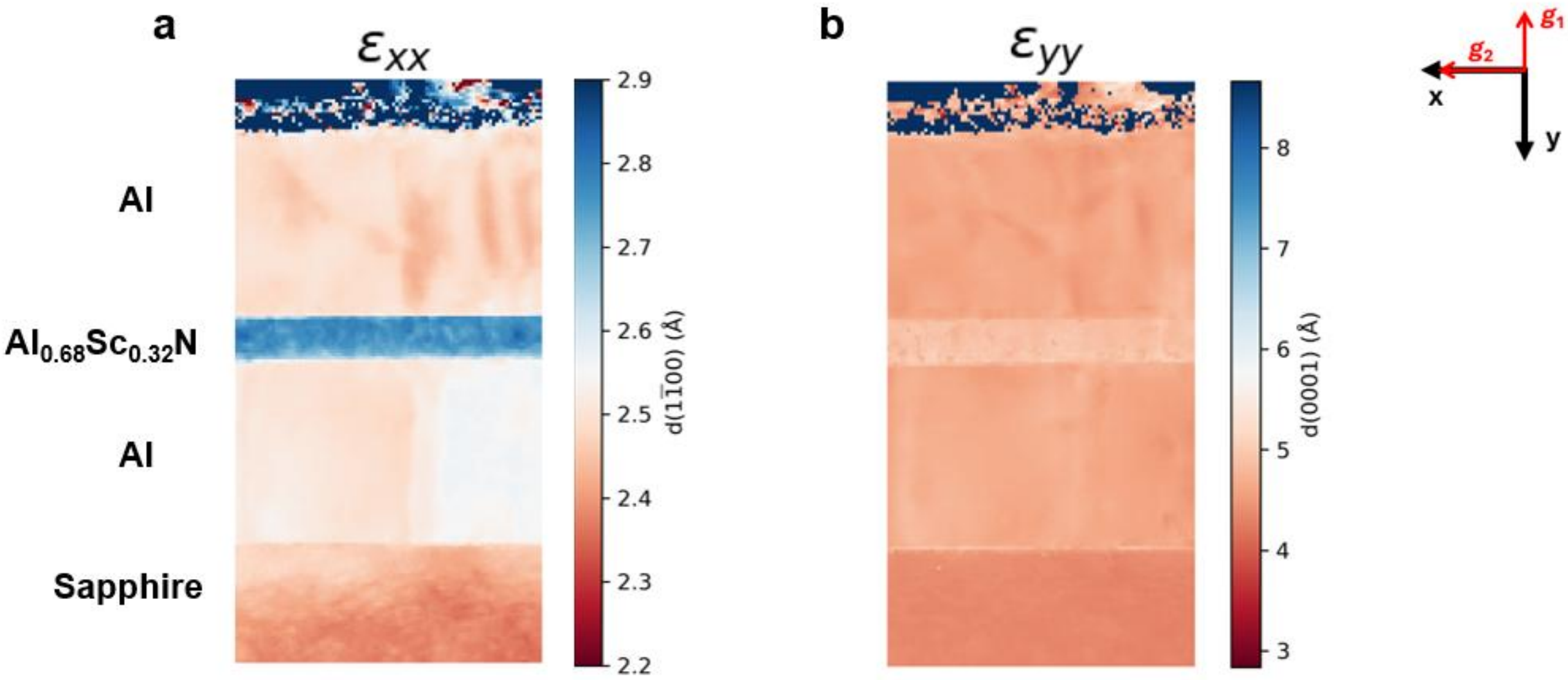

**Figure S15.** Out-of-plane and in-plane mapping of sapphire/Al/$Al_{0.68}Sc_{0.32}N$/Al calculated from strain mapping along $[\mathbf{1\bar{1}00}]$ of sapphire. (a) In-plane parameter mapping of AlScN. (b) Calculated out-of-plane parameter of AlScN. The lattice parameter map for Al overestimates the plane spacings as the measurements are done with sapphire as a reference.

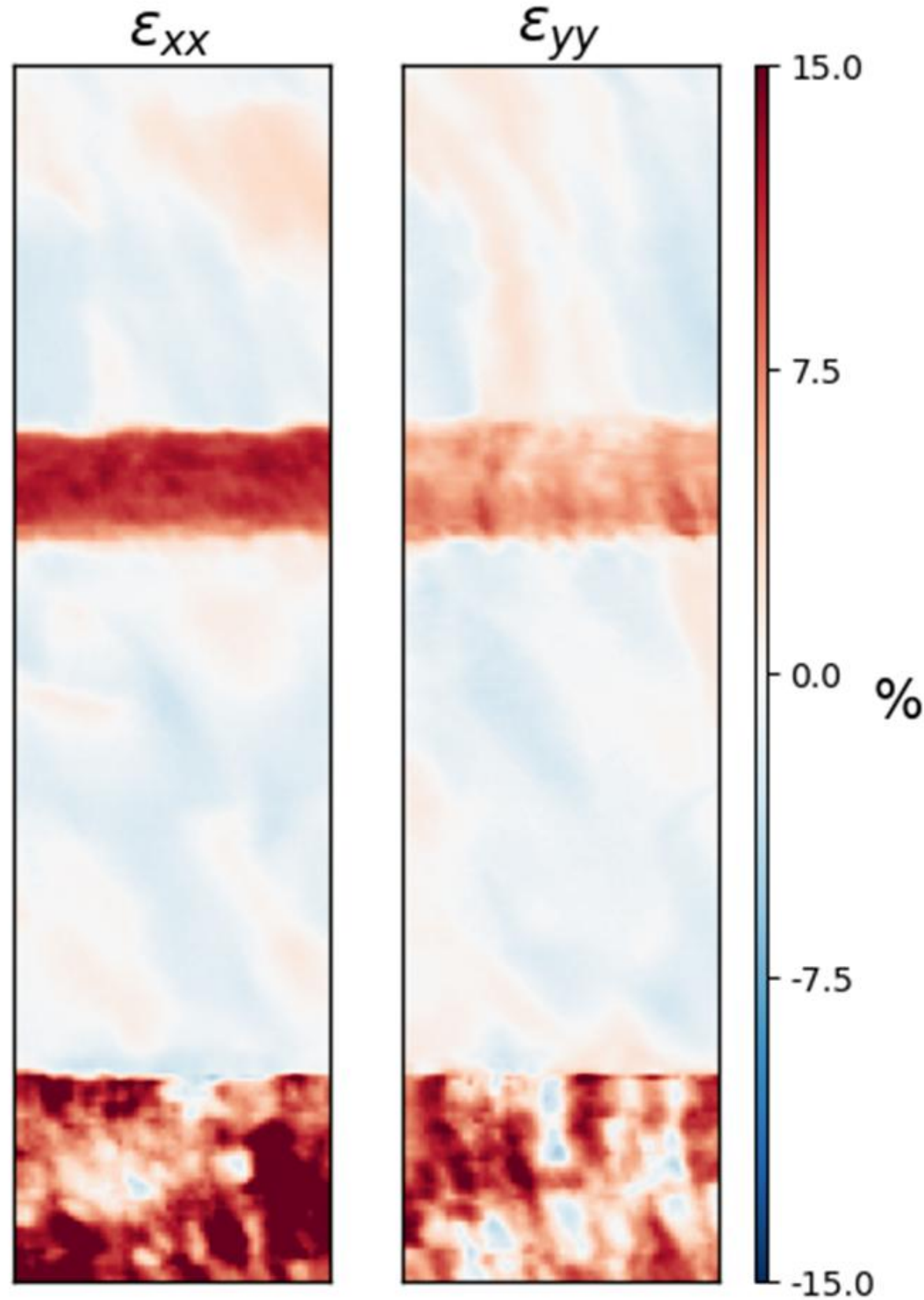

**Figure S16.** Strain mapping of sapphire/Al/$Al_{0.68}Sc_{0.32}N$/Al with Al and AlScN symmetrical reciprocal orientation